\documentclass[prb,showpacs,preprintnumbers,amsmath,amssymb,twocolumn,superscriptaddress]{revtex4-1}

\usepackage{graphicx}
\usepackage{dcolumn}
\usepackage{bm}
\usepackage{color}
\usepackage{natbib}

\usepackage{amsmath}	
\usepackage{amsfonts}

\newcommand{\vecc}[1]{\mbox{\boldmath $#1$}}

\begin{document}

\title{
Non-thermodynamic nature of the orbital angular momentum in neutral fermionic superfluids
}

\author{Yasuhiro Tada}
\affiliation{Institute for Solid State Physics, University of Tokyo, Kashiwa 277-8581, Japan}
\affiliation{Max Planck Institute for the Physics of Complex Systems, 
N{\"o}thnitzer Str. 38, 01187 Dresden, Germany}

\begin{abstract}
We discuss the orbital angular momentum (OAM) and the edge mass current in neutral fermionic superfluids with broken time
reversal symmetry.
Recent mean field studies imply that total OAM of a uniform superfluid depends on boundary conditions and is not a thermodynamic
quantity.
We point out that this does not conflict with thermodynamics,
because there is no intensive external field conjugate to OAM with which a uniform superfluid is stable in 
the thermodynamic limit,
in sharp contrast to the orbital magnetization in a non-superfluid system. 
We establish a simple physical picture for the sensitivity of OAM to boundaries by introducing the notion of 
``unpaired fermions'' and ``fermionic Landau criterion'' within a mean field description.
In order to go beyond the mean field approximation, we perform
a density matrix renormalization group calculation and conclude that the mean field understanding is essentially
correct.
\end{abstract}

\pacs{Valid PACS appear here}

\maketitle

\section{introduction}
\label{sec:intro}

The study of orbital angular momentum (OAM) in a superfluid with broken time reversal symmetry has a long history
since the discovery of $^3$He-A phase~\cite{book:Vollhart1990,book:Volovik2003,
pap:Leggett1975,book:Leggett2006,lecture:Leggett2012,pap:Mizushima2016}.
In the $p$-wave $^3$He-A phase, each Cooper pair is expected to carry OAM $\nu=1$, resulting in a bulk OAM proportional
to the total number of fermions $N$ in the system.
However, the discussions on the magnitude of the total OAM has been controversial and called ``intrinsic 
angular momentum paradox''.
At an intuitive level, the OAM is estimated as $L_z=\nu\times N/2$ just by counting the total number of Cooper pairs,
while OAM could also be estimated as $L_z=\nu\times (N/2)\times (\Delta/\varepsilon_F)$ since
the fermions only around the Fermi energy $\varepsilon_F$ would be relevant to physical quantities.
$\Delta$ is the gap amplitude of the superfluidity.
Both of the physical estimations seem to be reasonable, and detailed theoretical calculations predicted various values of the spontaneous OAM
corresponding to these physical pictures~\cite{book:Vollhart1990,book:Volovik2003,
pap:Leggett1975,book:Leggett2006,lecture:Leggett2012,pap:Mizushima2016,
pap:Ishikawa1977,pap:Ishikawa1980,pap:McClure1979,pap:Mermin1977,pap:Mermin1980,pap:Volovik1995,
pap:Kita1996,pap:Kita1998,pap:Goryo1998,pap:Anderson1961,pap:Volovik1975,pap:Volovik1976,pap:Cross1975,pap:Cross1977,pap:Combescot1978,pap:Leggett1978,pap:Hall1985,pap:Liu1985}.

Recently, this problem attracts renewed interest, partly
because the chiral superfluidity like $^3$He-A state is a prototypical example of topological superconductivity/superfluidity
~\cite{pap:Mizushima2016,pap:Stone2008,pap:Sauls2011,pap:Huang2014,pap:Huang2015,pap:Tada2015PRL,pap:Volovik2015,pap:Ojanen2016,pap:Scaffidi2015,pap:Tsuruta2015,pap:Suzuki2016}.
In such a chiral topological state, the edge mass current $J_{\rm edge}$ flows along a sample boundary,
which leads to OAM $L_z= 2J_{\rm edge}\times V$ where $V$ is the sample volume.
Interestingly, it has been proposed that the OAM is related to non-dissipative transport phenomena in two-dimensional
chiral superfluids~\cite{pap:Nomura2012,pap:Gromov2015,pap:Read2009,pap:Read2011,pap:Bradlyn2012,pap:Hoyos2014,pap:Shitade2014};
the thermal Hall conductivity $\kappa_{H}$ is given by temperature derivative of OAM, 
and the quantization of $\kappa_{H}$ at low temperature which is the hallmark of a chiral superfluid as a symmetry protected topological phase
is attributed to the quantized value of the edge mode contribution to OAM~\cite{pap:Nomura2012,pap:Gromov2015}.
Similarly, the Hall viscosity $\eta_H$ at zero temperature which is considered as an intrinsic non-dissipative transport quantity in two-dimensions
is shown to be proportional to OAM per fermion at and is therefore quantized in chiral superfluids~\cite{pap:Read2009,pap:Read2011,pap:Bradlyn2012,pap:Hoyos2014,pap:Shitade2014}.
The mass current Hall conductivity is also related to the OAM via $\eta_H$~\cite{pap:Hoyos2014}.
Since $\kappa_{H}$ and $\eta_H$ are considered as topological intrinsic quantities,
their connections to OAM would imply that $L_z$ is independent of details of the system such as the gap amplitude $\Delta$
and the Fermi energy $\varepsilon_F$.

However, surprisingly,
there have been various mean field calculations which show
that the spontaneous OAM in a chiral superfluid does depend on boundary conditions of a system
and is not an intrinsic quantity
~\cite{pap:Sauls2011,pap:Huang2014,pap:Huang2015,pap:Tada2015PRL,pap:Volovik2015,pap:Ojanen2016,pap:Scaffidi2015,pap:Tada2015,pap:Nagato2011,pap:Lederer2014,pap:Bouhon2014}. 
This is, on one hand, quite counter-intuitive, since it has been widely regarded that
OAM is a thermodynamic quantity and should be independent of 
non-thermodynamic details such as boundary conditions and shapes of a system.
This can be inferred from the well-known formula for the total OAM,
\begin{align}
L_z=-\frac{\partial F({\Omega_z})}{\partial \Omega_z},
\label{eq:F_OAM}
\end{align}
where $F(\Omega_z)$ is the free energy in the rotating frame with the angular velocity $\Omega_z$ along $z$-axis~\cite{book:LandauLifshitz}.
On the other hand, the sensitivity of OAM to boundaries would be rather natural, 
since the origin of spontaneous OAM in a chiral superfluid is mainly the edge mass current and
such a current could be influenced by details of boundaries along which the current flows. 
For example, we may expect that edge mass current depends on roughness of the sample surface.
Indeed, this was shown to occur within mean field calculations~\cite{pap:Sauls2011,pap:Tada2015,pap:Nagato2011,pap:Lederer2014}.
It was also demonstrated that 
directions of edge mass current can be even reversed depending on sample shapes~\cite{pap:Bouhon2014}.
Similar reversal of edge mass current is possible at a domain boundary between
superfluids with positive and negative chiralities~\cite{pap:Tsutsumi2014,pap:Tada2018}.
All the contributions to the energy from boundary conditions and system shapes
themselves are at most proportional to the surface area
in a system with short range interactions.

One can compare the OAM in superfluids with spontaneous orbital magnetization (OM) in non-superfluid systems,
and will find a qualitative difference between these two quantities.
Total spotaneous OM for a finite size system is simply proportional to total OAM at zero magnetic flux density, $M_z=-\mu_B L_z$, 
as an operator in an appropriately chosen gauge where $\mu_B$ is the Bohr magneton~\cite{com:OM_OAM}.
Although one might expect that the spontaneous OM is also a non-thermodynamic quantity which depends on boundaries or shapes,
it is a thermodynamic quantity and the well known formula,
\begin{align}
M_z=-\frac{\partial F(B_z)}{\partial B_z},
\label{eq:F_OM}
\end{align}
is a thermodynamic relation, where $B_z$ is the uniform magnetic flux density~\cite{com:OM_EM,pap:Banerjee1998}.
Indeed, these have been proved to be true in non-interacting systems~\cite{pap:Angelescu1975,
pap:Macris1988,pap:Briet2012}, and free energy density in interacting systems
have also been discussed extensively~\cite{book:Ruelle1999,book:Lieb2010}.
Therefore, 
the non-thermodynamic nature of spontaneous OAM discussed in the previous studies would be a characteristic property
only in superfluids.

Experimentally, direct observations of OAM and corresponding edge mass currents are challenging issues.
For example, there have been a limited number of experimental reports on the intrinsic angular momentum paradox in the $^3$He-A phase~\cite{pap:OAMexp1,pap:OAMexp2,proc:Ishikawa}, 
and the magnitude of the edge charge current in the candidate chiral $p$-wave 
superconductor Sr$_2$RuO$_4$ is extremely small compared with a theoretical estimation~\cite{pap:Hicks2010,pap:Scaffidi2015}.
If OAM and edge currents are boundary sensitive quantities, careful discussions will be required for an experimental detetction.

Although there have been many calculations of spontaneous OAM and corresponding edge mass current for concrete models of superfluids
within mean field approximations
~\cite{pap:Sauls2011,pap:Huang2014,pap:Huang2015,pap:Tada2015PRL,pap:Volovik2015,pap:Ojanen2016,pap:Scaffidi2015,pap:Tsuruta2015,pap:Suzuki2016,pap:Tada2015,pap:Nagato2011,pap:Lederer2014,pap:Bouhon2014}, 
a comprehensive understanding especially on connections to thermodynamics
has not been well established.
Therefore, it is desirable to develop a simple understanding on the physical reason why 
OAM in superfluids can depend on non-thermodynamic details, and obtain an intuitive picture. 
In this study, 
we point out that the OAM, especially the spontaneous OAM, in a superfluid is not a thermodynamic quantity by focusing on 
absence of a thermodynamic limit under rotation and roles of Hess-Fairbank effect under an artificial magnetic flux density.
Then, we establish a simple physical picture for the non-thermodynamic nature of the OAM within mean field descriptions,
where two important notions, ``unpaired fermions'' and ``fermionic Landau criterion'', are introduced.
In order to go beyond the mean field approximations,
we perform a non-perturbative numerical calculation by using the infinite density matrix renormalization group (iDMRG)~\cite{pap:DMRG0,pap:DMRG1,pap:DMRG2,pap:DMRG3,pap:TenPy,pap:TenPy2}.
It is concluded that the mean field understanding is essentially correct.

\section{non-thermodynamics of OAM in neutral superfluid}
\label{sec:TDL}
In this section, we discuss whether or not OAM (especially spontaneous OAM) in a uniform superfluid can be 
described within the standard thermodynamics.
Although this will be an elementaly discussion, to the best of our knowledge,
it has never been explicitly considered in the context of
the spontaneous OAM in chiral superfluids. 
This may be a reason for the controversial discussions on the OAM, and
therefore we will clarify some important points here.

\subsection{General discussion}

The extensive thermodynamic free energy $F^{\rm TD}_V$ for volume $V$ can be derived from the statistical mechanical free energy density in the thermodynamic limit $f_{\infty}$,
\begin{align}
F^{\rm TD}_V=V\times f_{\infty}.
\end{align}
If $F^{\rm TD}_V$ is well-defined in the presence of $\Omega_z\neq0$ or $B_z\neq0$,
Eq. \eqref{eq:F_OAM} or ~\eqref{eq:F_OM} becomes a thermodynamic relation.
Since a thermodynamic free energy is stable to non-extensive perturbations such as boundary conditions and shapes,
a physical extensive quantity obtained from $F^{\rm TD}_V$ should also be thermodynamic.
However, in general, it is a non-trivial problem whether or not a microscopic model has a well-defined thermodynamic limit
and thermodynamics can be applied to the system.
Indeed, the previous mean field studies may imply that Eqs. \eqref{eq:F_OAM} and ~\eqref{eq:F_OM} for a uniform superfluid
are not thermodynamic relations
~\cite{pap:Sauls2011,pap:Huang2014,pap:Huang2015,pap:Tada2015PRL,pap:Volovik2015,pap:Ojanen2016,pap:Scaffidi2015,pap:Tsuruta2015,pap:Suzuki2016,pap:Tada2015,pap:Nagato2011,pap:Lederer2014,pap:Bouhon2014}.

For a general system, we would naively expect that {\it an extensive quantity $M$ is
stable to non-extensive perturbations if and only if it is derived from a thermodynamic free energy.}
Indeed, if the thermodynamic free energy is obtained in the presence of a conjugate intensive field $h$ to $M$,
we have
\begin{align}
M^{\rm TD}_V(h)&=-\frac{\partial F^{\rm TD}_V(h)}{\partial h},
\end{align}
where $M^{\rm TD}_V$ is the thermodynamic value of the statistical mechanical quantity $M$
~\cite{com:F_derivative,pap:Lebowitz1968}.
Similarly, if the statistical mechanical expectation value $\langle M\rangle_V$ is robust to non-extensive perturbations,
the statistical mechanical free energy $F_V$ with the intensive conjugate field $h\neq0$ obtained as
\begin{align}
F_V(h)-F_V(0)&=-\int_0^h dh'\langle M\rangle_V (h')
\end{align}
will also be stable to the perturbations, if possible changes in $F_V(0)$ by the perturbations
are at most $o(V)$~\cite{com:F0}.
For such a stable $F_V(h)$, one would naively expect existence of a well defined thermodynamic limit.

An important subtle point in OAM is that the OAM operator itself is not extensive~\cite{pap:Cross1977,pap:Leggett1978,com:extensive_operator}.
For a system on $V\subset {\mathbb R}^3$, the OAM operator is defined as 
\begin{align}
L_z&=\int_V\psi^{\dagger}_{}[\vecc{r}\times \vecc{p}]_z\psi_{}d^3x=
\int _{V}[\vecc{r}\times \vecc{j}_{\rm }]_zd^3x,\\
\vecc{j}_{\rm }(\vecc{r})&=-i(\psi_{}^{\dagger}\nabla\psi_{}-\nabla\psi^{\dagger}_{}\psi_{})/2,
\end{align}
where $\vecc{p}_j=-i\nabla_j$ and $\vecc{j}_{\rm }$ is the mass current density.
Or equivalently, in the first quantization form,
\begin{align}
L_z&=\sum_{i=1}^N\vecc{r}_i\times \vecc{p}_i=\int _{V}[\vecc{r}\times \vecc{j}_{\rm }]_zd^3x,\\
\vecc{j}_{\rm }(\vecc{r})&=\sum_{i=1}^N\{\delta(\vecc{r}-\vecc{r}_i),\vecc{p}_i\}/2.
\end{align}
Although it is not trivial from these expressions, its expectation value at equilibrium
scales as $\langle L_z\rangle_V=O(V)$, if an equilibrium state is well-defined.
This is because $\vecc{j}_{\rm }(\vecc{r})$ is usually localized around the boundary of $V$ in such a state~\cite{pap:TadaKoma2016},
and the above expressions can be reduced to
\begin{align}
\langle L_z\rangle_V &\simeq \int_{\partial V} [\vecc{r}\times \vecc{J}_{\rm edge}]_zd^2x_{\parallel},\\
\vecc{J}_{\rm edge}(\vecc{r}_{\parallel})&=\int \langle\vecc{j}_{\rm }(\vecc{r})\rangle_V dx_{\perp},
\end{align}
where $\vecc{J}_{\rm edge}$ is the net edge current which are the integral of $\langle\vecc{j}_{\rm }\rangle_V$
over the perpendicular direction $x_{\perp}$ to the surface~\cite{com:J_def}.
Then, it is clear that $\langle L_z\rangle_V =O(V^{2/3})\times O(V^{1/3})=O(V)$.

We note that it is impossible to express the operator $L_z$ as a sum of local ``OAM density operator''
which is translationally symmetric~\cite{pap:Cross1977,pap:Leggett1978}. 
This can be understood as follows. Suppose that there exists a local OAM density operator of the form
$\vecc{l}(\vecc{r})=[\psi^{\dagger}(\vecc{r})\hat{\vecc{l}}(\nabla)\psi(\vecc{r})+({\rm h.c.})]$ where
$\hat{\vecc{l}}$ is independent of $\vecc{r}$.
We expand $\hat{l}_j=a_{1j}+a_{2jk}\nabla_k+\cdots$ where $a_{1j},a_{2jk},\cdots\in {\mathbb C}$.
Then it is easy to see that there is no solution for the 
commutation relation $[l_i(\vecc{r}),l_j(\vecc{r}')]=i\delta(\vecc{r}-\vecc{r}')\epsilon_{ijk}l_k(\vecc{r})$.
Therefore, the operator $\vecc{l}(\vecc{r})$ does not exist~\cite{com:j_operator,pap:Resta2013,pap:Resta2016}.
This is in sharp contrast to the familiar spin magnetization operator which can be equivalently expressed either
of the form $\int_Vd^3x \vecc{s}(\vecc{r})$ or $\int_Vd^3x [\vecc{r}\times (\nabla\times\vecc{s}(\vecc{r}))]$ with 
$\vecc{s}(\vecc{r})=\psi^{\dagger}\vecc{\sigma}\psi/2$.
These are consistent with our naive expectation that spin is an internal degrees of freedom which has the spatial position
independent generator of SU(2) symmetry, while an orbital motion is an spatially extended object and therefore
there is no local, translationally symmetric generator of SO(3) rotational symmetry.

In the following,
we will consider several theoretical setups in which OAM might possibly be obtained 
by derivative of a thermodynamic free energy; 
we introduce two kinds of external fields, a uniform rotation with or without additional confinement potentials
corresponding to Eq.\eqref{eq:F_OAM},
and an artificial constant magnetic flux density corresponding to Eq.\eqref{eq:F_OM}.
In each case, we explain that $F^{\rm TD}_V(h)$ is not obtained for a superfluid with a uniform density,
and OAM cannot be regarded as a thermodynamic quantity.

\subsection{System under rotation}

The robustness of $\langle M\rangle_V$ is guaranteed by the existence of a thermodynamic limit of $f_V(h)=F_V(h)/V$
under the intensive conjugate field $h$ for fixed $N/V$, which is often implicitly assumed in condensed matter physics.
Although this assumption is indeed satisfied in many systems, 
there are several important exceptions and a system of particles under a uniform rotation, for example, confined in a cylinder of radius $R$ is the case
~\cite{pap:Widom1968,pap:Lebowitz1968,com:polarization,pap:Avron1977,pap:Blanc2002}.
This is simply because the centrifugal potential $V_{\rm cen}\propto -(\Omega_z r_{\perp})^2$ ($r_{\perp}$ is the in-plane distance from the rotation axis) 
will push the particles onto the boundary of a system
and velocities of those particles $\Omega_z R$ will be infinitely fast when the system size $R$ 
becomes $R\rightarrow\infty$ with keeping $\Omega_z\neq0$.
Indeed, it is easy to show that such a system described by Schr{\"o}dinger Hamiltonian with a stable, short-range interaction
does not have stability of Hamiltonian of the second kind,
i.e. $^{\neg}(H>{\rm const}\times V)$, and therefore does not have a thermodynamic limit
when the particles are confined in a rigid wall container~\cite{pap:Widom1968}.
For example, we consider 
variational single particle wavefunctions 
\begin{align}
\psi_j(r_j,\theta_j,z_j)\simeq e^{il_j \theta_j}\tilde{\psi}_j(r_j),
\end{align}
where $l_j=\Omega_z/\Omega_j$ with $\Omega_j=1/mR_j^2$ and the particle mass $m$. 
$\tilde{\psi}_j(r_j)$ is localized at $r_j=R_j=O(R)$ and satisfies a given boundary condition at $r_j=R$.
Although we only consider particular values of $\Omega_z$ for which $\psi_j$ is consistent with a given boundary condition
on $\theta_j$, there are many such $\Omega_z$s when $\Omega_0=1/mR^2$ is sufficiently small.
We then construct an anti-symmmetric many-body wavefunction $\Psi_{\rm var}$ for $N$ particles from these single particle wavefunctions,
which obviously gives $\langle \Psi_{\rm var}|H|\Psi_{\rm var}\rangle \sim -I_0\Omega_z^2/2=O(V^{5/3})$ with $I_0=mNR^2$ in the leading order.
It is noted that the expectation value of a stable, short range interaction term in the Hamiltonian is at most $O(V)$,
and therefore it is irrelevant here.

For such a system to have a thermodynamic limit, one needs to keep $\Omega_z R=O(1)$ at a fixed value when taking $R\rightarrow\infty$,
which means that the angular velocity $\Omega_z$ is no longer an intensive field conjugate to OAM
~\cite{com:relativistic,pap:Davies1996,pap:Duffy2003,pap:Maeda2016}.
Therefore, Eq.~\eqref{eq:F_OAM} is an equation which holds only in a small size system and is not a thermodynamic equation.
In fact, for such a case, the coupling term in the free energy is $-\Omega_z\langle L_z\rangle=O(R^{-1})O(V)=O(V^{2/3})$ when OAM is $O(V)$,
and this energy gain is comparable with possible surface perturbations and therefore cannot guarantee thermodynamic nature of OAM.
Note that it could be considered that the state with a uniform density distribution~\cite{pap:Leggett1970,pap:Leggett1998}
is metastable (or non-equilibrium) and is not a true equilibrium/ground state of a rotating three-dimensional system.

One may expect that the Hamiltonian can be made stable for fixed $\Omega_z=O(1)$ by use of
a mathematical trick of introducing an additional confinement potential such as $V_{\rm con}=C_nr_{\perp}^n$,
and taking an appropriate limit of $C_n\downarrow 0$.
This might reproduce the desired thermodynamic limit of the original free energy, 
if $V_{\rm con}$-dependence of the free energy density could be removed. 
In order to do so, we wish to take the limit,
$\lim_{C_n\downarrow0}\lim_{N\uparrow\infty}F_N(\Omega_z)/N$ for fixed $\Omega_z=O(1)$.
However, for such a Hamiltonian with a confinement potential, one needs to take a scaling limit of $C_n$ and $N$
to have an extensive free energy.
For example $n=2$, the kinetic Hamiltonian can be written as $(\vecc{p}-\vecc{A})^2/2m+m\tilde{\Omega}_z^2r_{\perp}^2/2$
where $\vecc{A}=m\Omega\times\vecc{r}$ and $\tilde{\Omega}_z^2=(C_2-\Omega_z^2)>0$ with an appropriately chosen $C_2$.
Although $C_2\downarrow0$ limit cannot be taken for keeping the Hamiltonian with a fixed $\Omega_z\neq0$ stable, 
a scaling limit is required similarly to other $n$ cases.
Indeed,
to have an extensive free energy $F_N\propto N$ of the fermion model, one needs to keep $N_2\tilde{\Omega}_z^2=O(1)$
and the free energy density will depend on the value of $N_2\tilde{\Omega}_z^2$, where $N_2$ is the total number of particles divided by
the length of the system in $z$-direction~\cite{pap:Yoshioka1992}. 
Therefore, we cannot take the desired thermodynamic limit where the free energy density is well-defined and $V_{\rm con}$-independent,
even if
the mathematical trick of an additional confinement potential is used.
The special case $\tilde{\Omega}_z=0$ for which the scaling limit is not required will be discussed in the next section.

In summary, we have seen that 
the angular velocity is not an intensive field conjugate to OAM.
Therefore, we conclude that, in general, OAM in a system with a uniform particle density is not a thermodynamic quantity.

\subsection{System with constant artificial magnetic flux}
\subsubsection{Non-interacting case}

In the previous section, we have seen that 
OAM in a uniform system is generally non-thermodynamic.
However, this does not necessarily mean that {\it spontaneous} OAM at zero external field $\Omega_z=0$ is 
sensitive to non-extensive perturbations.
As mentioned in Sec.~\ref{sec:intro}, OAM is equivalent to OM at zero field as an operator, and the latter is thermodynamic in metals and insulators.
If we consider a neutral system minimally coupled with an artificial gauge field $\vecc{A}=m\vecc{\Omega} \times \vecc{r}
\equiv \vecc{B}\times \vecc{r}/2$, 
the Hamiltonian is formally equivalent to the charged particles under a constant magnetic flux density in the symmetric gauge. 
Or equivalently, we can introduce $V_{\rm con}= m(\Omega_zr_{\perp})^2/2$ as a mathematical trick to make the Hamiltonian under rotation stable.
We note that a gauge invariant Hamiltonian $H$ now has the stability $H>{\rm const}\times V$ and
translational symmetry if combined with a suitable gauge transformation,
and we can take the desired limit $\lim_{V\uparrow\infty}F_V(B_z)/V$ for a fixed intensive field $B_z\neq 0$ in this case. 
If the ``charged'' system under uniform $B_z$ has a thermodynamic limit, 
spontaneous OAM in a neutral system obtained as the limit of $B_z\downarrow0$ (or $B_z\uparrow0$) will also be thermodynamic.
Indeed, this is true for non-superfluids and spontaneous OM/OAM obtained so will be a thermodynamic quantity
~\cite{pap:Angelescu1975,pap:Macris1988,pap:Briet2012,book:Ruelle1999,book:Lieb2010}.
We will give a brief explanation on the existence of the thermodynamic limit in Appendix~\ref{app:A}.

Similarly, one might expect that OAM in a neutral superfluid should be thermodynamic as well.
Unfortunately, this expectation is not correct, because for a Hamiltonian of a uniform superfluid under a constant artificial magnetic flux density $B_z$,
derivative of the free/ground state energy with respect to $B_z$ does not give the desired OM.
To see this, for simplicity,
let us consider a non-interacting fermionic Hamiltonian with a uniform $s$-wave gap function $\Delta_0$ as a U(1) symmetry breaking field,
$H=\int d^3x \psi^{\dagger}[(\vecc{p}-\vecc{A})^2/2m]\psi+\Delta_0\int d^3x \psi^{\dagger}_{\uparrow}\psi^{\dagger}_{\downarrow}+({\rm h.c.})$.
This Hamiltonian is not gauge invariant and does not have translation symmetry, and therefore 
existence of a thermodynamic limit in the presence of $B_z\neq0$ is not guaranteed.
Indeed, the induced current is given by $\langle \vecc{j}\rangle \simeq -\langle (\psi^{\dagger}\psi\rangle/m)\vecc{A}\propto \vecc{B}\times\vecc{r}$,
and OAM per volume and free energy density diverge at $V\rightarrow\infty$, which is unphysical~\cite{com:Bq}.
The reason for the divergence is very simple; 
physically, the dangerous behavior of $\langle \vecc{j}\rangle$ comes from the fact that a uniform real magnetic 
flux
density cannot be realized in superconductors because of
the Meissner effect where electromagnetic field is determined by the Maxwell equation. 
The introduction of the constant artificial $B$-field into the Hamiltonian corresponds to an implicit
assumption that there is no Meissner or Hess-Fairbank effect. 
Such an unphysical assumption results in a huge energy cost, leading
to the superextensive free energy $F_V$.

\begin{figure}[htbp]
\begin{tabular}{cc}
\begin{minipage}{0.5\hsize}
\begin{center}
\includegraphics[width=\hsize,height=0.6\hsize]{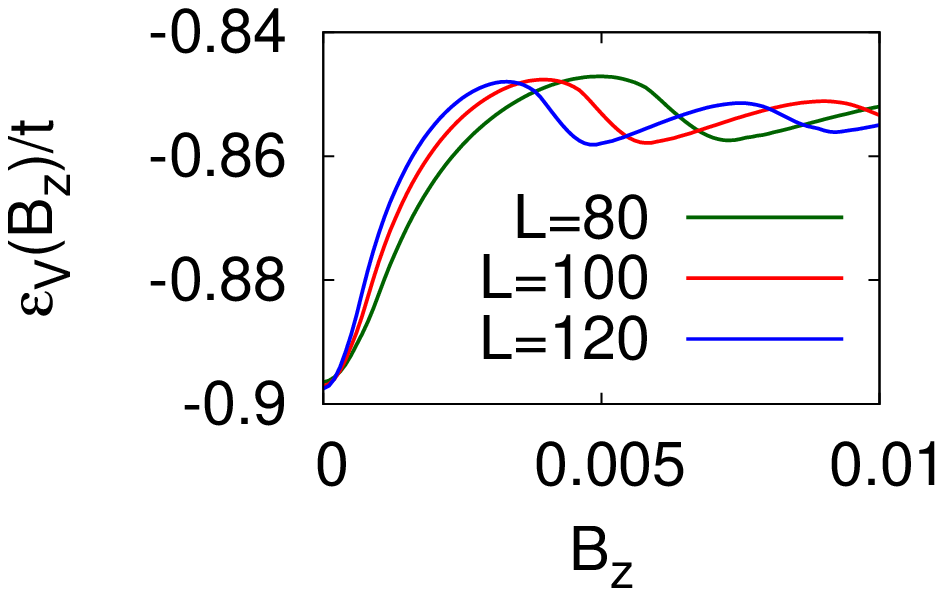}
\end{center}
\end{minipage}
\begin{minipage}{0.5\hsize}
\begin{center}
\includegraphics[width=\hsize,height=0.6\hsize]{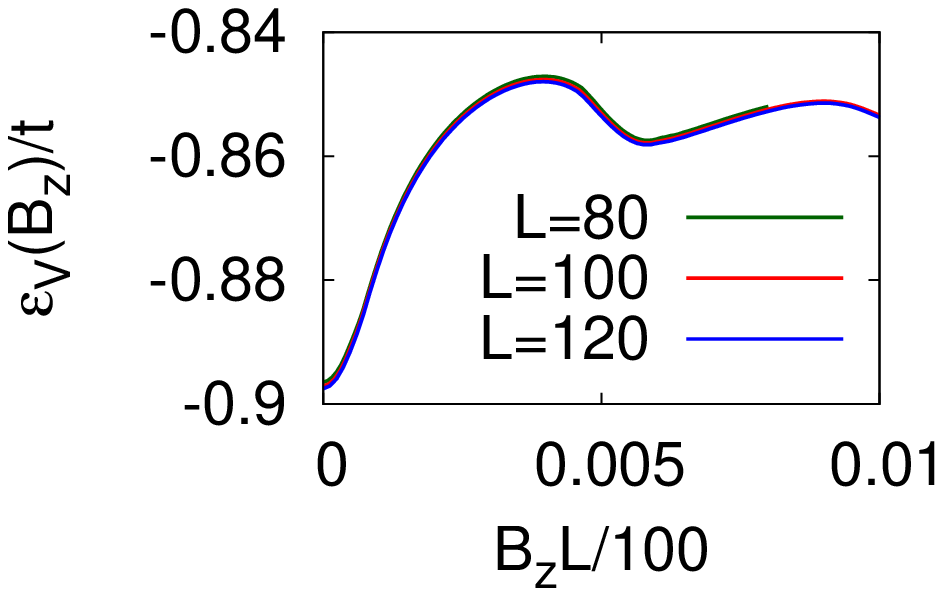}
\end{center}
\end{minipage}
\end{tabular}
\caption{(Left panel) The ground state energy density at $\Delta_0=0.6t$ for different system sizes.
The dimensionless magnetic flux density is $B_z=\phi/2\pi$ where $\phi$ is the flux per plaquette of the square lattice.
(Right panel) The ground state energy density in the scaling plot.}
\label{fig:E_scale}
\end{figure}

For a system defined on a cube with the volume $V=L^3$, the superextensive $f_V(B)$ behaves as $f_V(B)=\tilde{f}(LB)$
where $\tilde{f}$ is a scaling function which is nearly independent of $V$.
This is because the additional energy density due to the $B$-field is $B_z\langle M_z\rangle_V/L^3\sim B_z^2\int d^3x (x^2+y^2)/L^3\sim (B_zL)^2$,
where we have used $\langle \vecc{j}\rangle_V \sim\vecc{A}= \vecc{B}\times \vecc{r}/2$.
This scaling behavior is also seen in lattice models.
Here, we consider a two-dimensional square lattice of $V=L^2$ with open boundaries and spinful fermions at half filling with a fixed uniform $s$-wave gap function as a simple example,
\begin{align}
H=\sum_{\langle i,j\rangle,\sigma}-te^{iA_{ij}}c_{i\sigma}^{\dagger}c_{j\sigma}
+\sum_{i}[\Delta_0c_{i\uparrow}^{\dagger}c_{i\downarrow}^{\dagger}+({\rm h.c.})]. 
\end{align}
The vector potential $A_{ij}$ describes a constant magnetic flux density $B_z$ in the symmetric gauge.
We show the ground state energy density $\epsilon_V(B_z)=\langle H\rangle_V (B_z)/V$ in Fig.\ref{fig:E_scale}.
$\epsilon_V(B_z)$ has a strong size dependence, and all the data collapse into a single curve in the scaling plot,
$\epsilon_V(B_z)=\tilde{\epsilon}(B_zL)$. 
For small $B_zL$, the scaling function behaves as $\tilde{\epsilon}(B_zL)\sim (B_zL)^2$ as expected from the above discussion for
a continuum system.
The scaling behavior clearly shows the absence of the thermodynamic limit at a fixed $B_z\neq0$.

Finally, we note that, in a realistic charged superconductor with electromagnetic fields described by the Maxwell equation, 
the Meissner effect arises from the combination of (i) a response of electrons to a given vector potential and (ii) dynamics of electromagnetic fields 
in presence of a given electron current. 
Note that the two contributions (i) and (ii) to the total edge current are spatially separated with different length scales, 
the coherence length for (i) and the penetration depth for (ii). 
In the present study, the artificial vector field is a given fixed field and we do not consider its dynamics. 
On the other hand, the current density as the response (i) should be essentially proportional to the given artificial vector potential in superfluids, 
which we call ``strong diamagnetic response''. 
It is noted that the induced current density is not necessarily localized at a boundary for a general vector field.

\subsubsection{Interacting case}

Now we turn to interacting systems under a physically reasonable assumption.
In an interacting system, there are two ways for describing a uniform superfluid, one with explicit U(1) symmetry breaking and the other with conserved U(1) symmetry
~\cite{com:U(1)symmetry,pap:Shimizu2001}.
In both descriptions, the strong diamagnetic response, i.e. Meissner or Hess-Fairbank effect is a necessary condition for the superfluidity.
Then, the assumption is that there exists the strong diamagnetic response where the current density
is essentially proportional to a given vector potential, 
when (i) a uniform U(1) symmetry breaking field $\Delta_0$ is introduced into the Hamiltonian, or
(ii) there is a uniform long range order of the particle number U(1) symmetry in the absence of $\Delta_0$.
The uniform long range order is defined as 
\begin{align}
\sigma_{\rm C}&=\lim_{V\uparrow\infty}\sqrt{\langle m_{\rm C}^2\rangle_V},\\
m_{\rm C}&=\frac{1}{V}\int_V d^3x[\psi_{\uparrow}d(-i\nabla)\psi_{\downarrow}+({\rm h.c})],
\end{align}
where $m_{\rm C}$ is the uniform Cooper pair order parameter 
with a form factor $d=1$ for $s$-wave, $d\sim (p_x+ip_y)$ with $p_{j}=-i\partial_j$ for chiral $p$-wave, and so on.
This assumption is widely accepted and 
guarantees presence of the Meissner or Hess-Fairbank effect. 
The real Meissner effect is realized when combined with the Maxwell equation, but the present artificial vector potential
is simply a given field, $|\vecc{A}|\propto |\vecc{r}|$.
It should be noted that two approaches corresponding to (i) and (ii)
are equivalent for evaluating gauge invariant quantities per volume, 
if there exist the thermodynamic limits for both cases.
Corresponding to the two approaches, we consider interacting superfluids in two ways in the following.

In the first scheme (i), we describe a  superfluid as a global U(1) symmetry broken state,
where an introduced U(1) symmetry breaking field $\Delta_0$ must be turned off
after the thermodynamic limit is taken.
In this case, from the above assumption, there exists a contribution to the current from the symmetry breaking field and
$\langle \vecc{j}\rangle$ should contain a term essentially proportional to $\vecc{A}$.
Similarly to the non-interacting case, this leads to an unphysical divergence of the free energy density,
since the symmetry breaking field should be kept constant when the thermodynamic limit is taken.
Physically, the Hamiltonian with a constant $B_z$-field would correspond to a vortex state or a normal (non-superfluid) state.
Therefore, if necessary, 
one might need to introduce new corresponding symmetry breaking field which is different from the uniform field $\Delta_0$.

In the other scheme (ii), we do not introduce a symmetry breaking field into the Hamiltonian, 
and the U(1) symmetry is strictly kept in both finite and infinite volume systems, although we assume that
the system has the uniform U(1) long range order in the absence of $B$-field.
In this case, there exists the thermodynamic limit of the free energy density under the uniform artificial $B$-field,
since the Hamiltonian is translationally invariant if combined with an appropriate gauge transformation. 
(See Appendix~\ref{app:A} for a brief discussion.)
This means that there is no U(1) long range order, since if it were there, the strong diamagnetic response 
will lead to a divergent free energy density.
Therefore, we conclude that a constant $B$-field will suppress the pre-existing uniform U(1) long range order.
The absence of the strong diamagnetic response can also be understood from the Bloch's theorem which excludes a macroscopic current 
in a ground state/equilibrium for a U(1) symmetric system~\cite{pap:TadaKoma2016}.
Besides, the suppression of the uniform long range order may be consistent with a variant of the Elitzur's theorem for a fixed gauge field configuration
derived in Ref.~\textcite{pap:TadaKoma2016},
according to which the long range order $\sigma_{\rm C}$ vanishes for almost all gauge field configurations. 
Although we have assumed that $\sigma_{\rm C}\neq0$ for the specially chosen gauge $\vecc{A}=0$ in the absence of $B$-field, 
a gauge field corresponding to the constant $B$-field
is no longer compatible with a uniformly Cooper paired state.
This is physically reasonable, since one would expect a vortex state or a normal (non-superfluid) state for Hamiltonian with the uniform $B$-field.
There might arise a new long range order such as a vortex state once a uniform $B$-field is introduced into the Hamiltonian, 
or any long range order of U(1) symmetry might get suppressed.

It is important to realize that the state at $B_z=0$ and that at $B_z\neq0$ are physically different states in distinct ``phases'',
because only the former has the uniform U(1) long range order.
The quantity we are interested in this study is $l_0=\lim_{\Delta_0\rightarrow0}\lim_{V\rightarrow\infty}\langle L_z\rangle_V(B_z=0)/V
=\lim_{\Delta_0\rightarrow0}\lim_{V\rightarrow\infty}\partial f_V(B_z=0,\Delta_0)/\partial B_z$,
while the latter state gives different quantities 
$l_{\pm}= -\partial f_{\infty}(B_z\rightarrow\pm0)/\partial B_{z\pm}$ 
~\cite{com:RashbaSF,pap:SatoFujimoto2016}.
As already mentioned, $l_0$ and $l_{\pm}$ are expectation values of $L_z$ at different states.
Although $l_{\pm}$ is a thermodynamic quantity by definition,
$l_0$ is not directly related to $f_{\infty}$ and is not thermodynamic in this sense.
We note that $f_V(B_z, \Delta_0)$ for general $(B_z, \Delta_0)$ contains a scaling term $\tilde{f}(B_zL,\Delta_0)$
and diverges as $V\rightarrow\infty$ except for $(B_z=0, \Delta_0)$ or $(B_z, \Delta_0=0)$.
It is also noted that the discussions based on the two schemes (i) and (ii) for describing a superfluid
are consistent, as expected.

In summary,
we have seen that thermodynamic limits of the free energy density do not exist 
in several theoretical setups which could seemingly realize the desired uniform superfluid state.
As a result,
the OAM in a neutral superfluid is not a thermodynamic quantity
at some value of external (rotation/artificial flux) fields,
although it is usually extensive and seemingly thermodynamic.
Therefore, it can depend on non-thermodynamic details such as boundary conditions.
In the next section,
we will demostrate a physical picture on how OAM is affected by non-extensive perturbations.

\section{mean field description of fragile OAM}
\label{sec:MF}
\subsection{Unpaired fermions and fermionic Landau criterion}
In the previous section, we have explained that spontaneous OAM of a neutral superfluid is not related to thermodynamic free energy.
Then, it is important to develop a physical understanding on behaviors of OAM in the presence of 
non-extensive perturbations.
In this section, we discuss a mean field understanding at zero external rotation or $B$-field which has potential
applicability to a large class of neutral superfluids. 
This part is based on the recent progress~\cite{pap:Huang2015,pap:Tada2015PRL,pap:Volovik2015,pap:Ojanen2016,pap:Tada2018}, and here we establish an intuitively clear picture for the
seemingly non-trivial sensitivity of OAM.

In order to demonstrate the essential physics, we consider a two-dimensional $d+id$-wave superfluid confined by a rotationally
symmetric potential $V_{\rm con}$ as a simple example.
We mainly focus on the weak coupling BCS states where edge states are topological and gapless.
The argument can also apply to rotationally asymmetric systems and in principle to the strong coupling BEC states
with some modifications, when OAM is determined by an edge mass current.
The mean field Hamiltonian with the rotationally symmetric confinement potential reads,
\begin{align}
H&=\int d^2x \psi_{\sigma}^{\dagger}\Bigl( \frac{\vecc{p}^2}{2m}-\mu+V_{\rm con}\Bigr)\psi_{\sigma}\nonumber\\
&\quad +\Delta_0 k_F^{-2}\int d^2x \psi^{\dagger}_{\uparrow} (p_x+ip_y)^2\psi^{\dagger}_{\downarrow} +({\rm h.c.}),
\label{eq:Hd}
\end{align}
where $p_j=-i\nabla_j$.
$k_F$ is the Fermi momentum in the normal state, and $\Delta_0$ is the symmetry breaking field.
In this section, the confinement potential is zero in the bulk of the system, $V_{\rm con}(r\ll R)=0$, and infinitely large outside of the system, 
$V_{\rm con}(r\gg R)=\infty$, where $R$ is the system radius.
We expand the field operator as $\psi_{\sigma}(r,\theta)=\sum_{ml}c_{ml\sigma}\phi_{ml}(r,\theta)$ by using 
the single particle eigenfunctions of
$[\vecc{p}^2/2m-\mu+V_{\rm con}]$.
The Hamiltonian is rewritten into a Bogoliubov-de Gennes form,
\begin{align}
H&=\sum_{m,l}
\left[
\begin{array}{c}
c^{\dagger}_{m,l+2,\uparrow}\\
c_{m,-l,\downarrow}
\end{array}\right]^T
\left(\hat{H}_{\rm BdG}^{(l)}\right)_{mm'}
\left[
\begin{array}{c}
c_{m',l+2,\uparrow}\\
c^{\dagger}_{m',-l,\downarrow}
\end{array}\right].
\end{align}
We first consider a smooth confinement potential 
$V_{\rm con}(r)$ which increases smoothly 
around $r\sim R$ with a length scale $\xi_{\rm con}$ which satisfies $\xi_{\Delta}=v_F/\Delta_0\ll \xi_{\rm con}
\ll R$. 
The total OAM at $T=0$ is easily calculated as~\cite{pap:Tada2015PRL,pap:Prem2017}
\begin{align}
\langle L_z\rangle_V &= 2\times \frac{\langle N\rangle_V}{2}-\frac{1}{2}\sum_l (l+1)\eta_l,\\
\eta_l&=\sum_n {\rm sgn}\varepsilon_n{(l)},
\label{eq:Lz_eta}
\end{align}
where $\eta_l$ is the spectral asymmetry of the BdG Hamiltonian $H_{\rm BdG}^{(l)}$
for which eigenvalues are $\{\varepsilon_{n}{(l)}\}$.
Within the semi-classical approximation which can be valified for $\Delta_0\ll \varepsilon_F$,
OAM at zero temperature is reduced to
\begin{align}
\langle L_z\rangle_V &\simeq 2\times \frac{\langle N\rangle_V}{2}-\frac{1}{2}\sum_{j=1,2} \bigl(Rk_{Fj} \bigr)^2.
\label{eq:semi-cl}
\end{align}
Here, we have introduced Fermi wavenumbers of the two one-dimensional edge modes, $k_{F1}=-k_{F2} (k_{F1}\leq k_{F2})$,
which are defined as $Rk_{Fj}=l_j$ with the vanishing eigenvalues of the edge modes, $\varepsilon_{n}(l_j)\simeq 0$. 
Since $\xi_{\Delta}=v_F/\Delta_0\ll \xi_{\rm con}$, within the semi-classical approximation, 
the confinement potential can be treated as a constant energy shift $\varepsilon_F
\rightarrow \varepsilon_F'=\varepsilon_F-V_{\rm con}(r=R)$
for the edge modes localized in the relatively low particle density region $|r-R|\lesssim \xi_{\rm con}$. 
Then the edge mode Fermi wavenumber is given by $k_{F1,2}=\pm k_F'/\sqrt{2}$~\cite{pap:Tada2015PRL}.
When the potential is so smooth with large $\xi_{\rm con}$ that
the particle density is vanishing around $r\simeq R$ for which $k_F'\simeq0$, Eq.\eqref{eq:semi-cl} gives 
the ``full" value $\langle L_z\rangle_V=\langle N\rangle_V$.

Now we deform the confinement potential by decreasing $\xi_{\rm con}$ so that $\xi_{\rm con}\ll \xi_{\Delta}\ll R$ is now satisfied.
Note that this is a microscopic deformation of $V_{\rm con}$ in a length scale much smaller than $R$, and $V_{\rm con}$ remains unchanged in the length scale $O(R)$.
The modified potential gives a sharp confinement, $V_{\rm con}(r<R)=0$ and $V_{\rm con}(r\geq R)=\infty$, in the limit of $\xi_{\rm con}\rightarrow0$,
and the edge mode Fermi wavenumbers are $k_{F1,2}=\pm k_F/\sqrt{2}$, resulting in $\langle L_z\rangle_V=0$ within the semi-classical approximation.
We show numerical calculations of the OAM for the sharp confinement in Appendix~\ref{app:B}.

\begin{figure}[htbp]
\begin{center}
\includegraphics[width=\hsize,height=0.3\hsize]{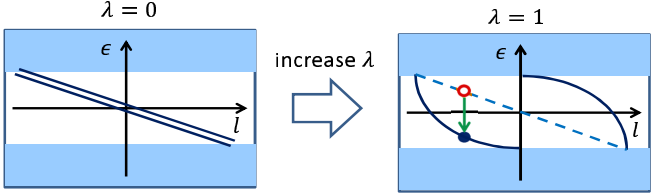}
\end{center}
\caption{Schematic picture of the spectrum of a chiral $d$-wave superfluid for the smooth confinement potential (left)
and sharp confinement potential (right). $\lambda$ characterizes the potential shape.
Some eigenvalues changes the sign as indicated by the green arrow.}
\label{fig:spec_flow}
\end{figure}

Although this $V_{\rm con}$-dependence of OAM seems curious at first sight,
the physical reason is simple as discussed below.
When we modify $V_{\rm con}$ which is parametrized by $0\leq\lambda\leq1$ ($\lambda=0$ for the smooth potential and
$\lambda=1$ for the sharp potential), some eigenvalues of the edge modes change their signs because $k_{F1,2}(\lambda=0)=0$ and $k_{F1,2}(\lambda=1)=\pm k_F/\sqrt{2}$,
as seen in Fig.\ref{fig:spec_flow}.
The ground state wavefunction satisfies $b^{\dagger}_{nl}|{\rm GS}(\lambda=0)\rangle=0$ when $\varepsilon_n(l;\lambda)<0$,
while $b^{}_{nl}|{\rm GS}(\lambda=1)\rangle=0$ when $\varepsilon_n(l;\lambda)>0$, 
where $b_{nl}$ is the $\lambda$-dependent annihilation operator of the eigen-mode of $\hat{H}_{\rm BdG}$ with $\varepsilon_{n}{(l;\lambda)}$.
Let us focus on an eigenvalue $\varepsilon_{n}{(l)}$ of the first edge mode ($j=1$) which is originally positive for the smooth potential and
becomes negative for the sharp potential at a critical $\lambda=\lambda_{nl}^c$. 
Because the ground state is characterized by $b_{nl}|{\rm GS}\rangle=0$ or $b_{nl}^{\dagger}|{\rm GS}\rangle=0$ depending on the sign of $\varepsilon_n(l)$,
there will be level crossing between the ground state and an excited state at $\lambda=\lambda_{nl}^c$.
If we denote the corresponding eigenstates as $|\Psi_0\rangle=|{\rm GS}\rangle$ and $|\Psi_1\rangle$,
these two states are related as $|\Psi_1\rangle\sim b_{nl}^{\dagger}|\Psi_0\rangle$. 
\begin{figure}[htbp]
\vspace{20pt}
\begin{center}
\includegraphics[width=0.5\hsize,height=0.3\hsize]{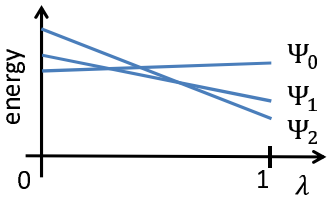}
\end{center}
\caption{
Schematic picture of successive level crossings between the eigenstates.
}
\label{fig:level}
\end{figure} 
Although the fermions are fully paried up at $\lambda<\lambda_{nl}^c$, 
the Cooper pair is partially broken by the application of $b_{nl}^{\dagger}$-operator for $\lambda>\lambda_{nl}^c$.
Once a Cooper pair of the chiral $d$-wave state which is 
expressed as $c_{l+2,\uparrow}^{\dagger}c_{-l,\downarrow}^{\dagger}|0\rangle$ ($|0\rangle$ is the vacuum of $c$-fermions)
and carries OAM$=2$ is broken, one fermion will be removed from $|{\rm GS}\rangle$ and 
an unpaired single fermion mode will remain filled. 
By a careful analysis, it turns out that $c_{l+2,\uparrow}^{\dagger}$-fermion will remain 
while $c_{-l,\downarrow}^{\dagger}$-fermion
is removed from the ground state where $l<-1$ in the first branch ($j=1$) of the edge modes, 
which reduces OAM as $(l+2)+(-l)=2\rightarrow (l+2)\leq 0$~\cite{pap:Tada2015PRL}.

If we consider other eigenvalues, there will be a sequence of spectral flows of $\{\varepsilon_n(l)\}$ and corresponding successive level crossings between
the eigenstates as shown in Fig.~\ref{fig:level}.
Consequently, by increasing $\lambda$,
the original ground state wavefunction 
$|{\rm GS}(\lambda=0)\rangle={\mathcal N}
\otimes_l|{\rm GS}(\lambda=0),l\rangle={\mathcal N}\otimes_l\exp[\sum_{jj'}\tilde{c}_{j,l+2,\uparrow}^{\dagger}
F^{(l)}_{jj^{\prime}}
\tilde{c}_{j^{\prime},-l,\downarrow}^{\dagger}]|0\rangle$ 
(${\mathcal N}$ is a normalization constant) for the smooth potential at $\lambda=0$
is replaced with the new ground state wavefunction for the sharp potential at $\lambda=1$, 
\begin{align}
|{\rm GS}(\lambda=1),l\rangle&=\left(\prod_{j=1}^{n_{\uparrow}^{(l)}}
\tilde{c}_{j,l+2,\uparrow}^{\dagger}\right)
\left(\prod_{j=1}^{n_{\downarrow}^{(l)}}
\tilde{c}_{j,-l,\downarrow}^{\dagger}\right)\notag\\
&\times 
\exp \left(\sum_{j>n_{\uparrow}^{(l)}}\sum_{j^{\prime}>n_{\downarrow}^{(l)}}
\tilde{c}_{j,l+2,\uparrow}^{\dagger}
F^{(l)}_{jj^{\prime}}
\tilde{c}_{j^{\prime},-l,\downarrow}^{\dagger}\right)|0\rangle.
\label{eq:GS}
\end{align}
The parameters $n_{\uparrow,\downarrow}^{(l)}(\lambda),F^{(l)}_{jj^{\prime}}(\lambda)$ and the explicit form of the $\tilde{c}$-operators
can be calculated from diagonalization of $\hat{H}_{\rm BdG}(\lambda)$
~\cite{pap:Tada2015PRL,pap:Prem2017}.
For $\lambda=1$, we have $(n_{\uparrow}^{(l)},n_{\downarrow}^{(l)})=(1,0)$ for $l_1<l<-1$, $(n_{\uparrow}^{(l)},n_{\downarrow}^{(l)})=(0,1)$ for $-1<l<l_2$,
and $(n_{\uparrow}^{(l)},n_{\downarrow}^{(l)})=(0,0)$ otherwise, where $l_j=k_{Fj}R$.
Note that although the number of unpaired fermions induced by the change in $V_{\rm con}$, i.e. the total number of $\{n_{\uparrow,\downarrow}^{(l)}\}_l$
in Eq.~\eqref{eq:GS},
is only $O(R)$ and their contributions to
the ground state energy are negligibly small, 
they have large impacts on the edge mass current and consequently on OAM.

We believe that the depairing effect of the Cooper pairs and resulting reductions of edge mass currents 
are a universal mechanism in fermionic neutral superfluids,
although we have used the very simple model as an example for an illustrative demonstration.
For example, a chiral $p$-wave system can show spectral flow depending on system shapes,
and the pair breaking effect works in lattice models as well where continuous rotational symmetry is absent~\cite{pap:Huang2015,pap:Bouhon2014,pap:Tada2018}.

The above mechanism is analogous to the well known Landau criterion for bosonic superfluids where
the preformed superfluid is broken once a bosonic excitation energy $\varepsilon(\lambda)$ becomes negative as some parameter is varied,
leading to a new condensation of this boson mode.
Similarly, in the present fermion case,
the Cooper pair is broken once its excitation energy becomes negative,
and the resulting ground state is a state with broken Cooper pairs, i.e. unpaired fermions.
The essential difference is that Cooper pairs are broken only for the modes with sign changing eigenvalues in the fermion case.
We call this partial breaking of fermion superfluidity as ``fermionic Landau criterion".
The similarity between the fermionic Landau criterion and bosonic Landau criterion will be discussed in detail in the next section.
It should be noted that the fermionic Landau criterion is based on a given mean field Hamiltonian
where the gap function is simply given and therefore it does not necessarily hold in general interacting models.
In a realistic interacting model, it may be possible that the system goes into a completely different phase
as the parameter $\lambda$ is varied, if the energy cost due to unpaired fermions is $O(V)$.

\subsection{Superfluid under uniform linear flow}

In this section, we study a toy model for a superfluid under a uniform linear flow~\cite{book:Tinkham2004,pap:Bardeen1962}
to clarify the analogy between the fermionic Landau criterion and well-known bosonic Landau criterion.
This analogy is helpful to understand the physics in a comprehensive way.

\begin{figure}[htbp]
\begin{tabular}{lr}
\begin{minipage}{0.35\hsize}
\begin{center}
\includegraphics[width=\hsize,height=\hsize]{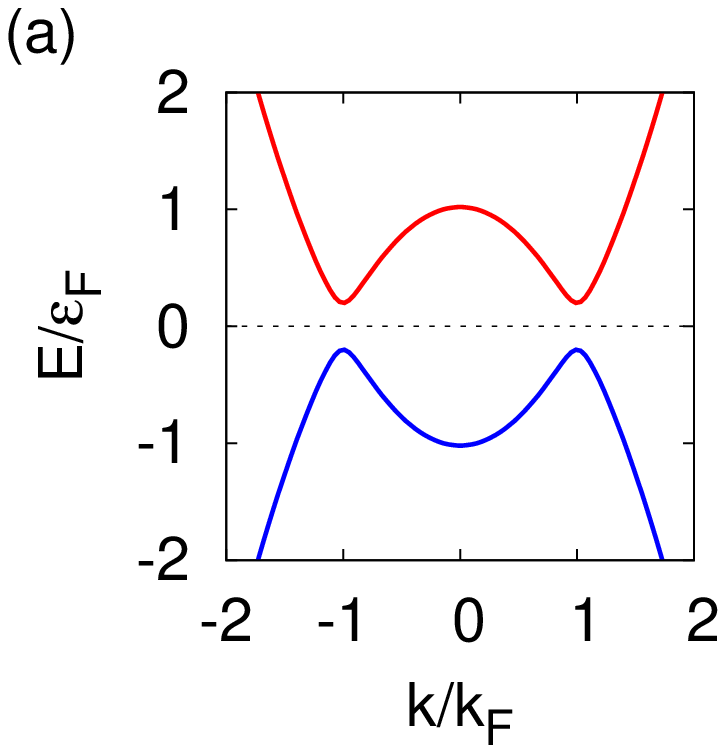}
\end{center}
\end{minipage}
\hspace{0.5cm}
\begin{minipage}{0.35\hsize}
\begin{center}
\includegraphics[width=\hsize,height=\hsize]{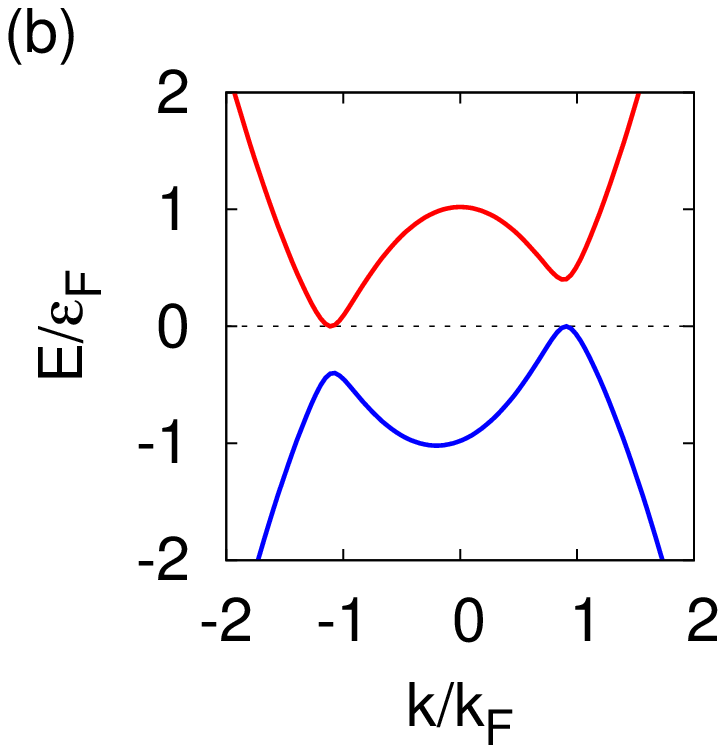}
\end{center}
\end{minipage}
\end{tabular}
\begin{tabular}{cc}
\begin{minipage}{0.35\hsize}
\begin{center}
\includegraphics[width=\hsize,height=\hsize]{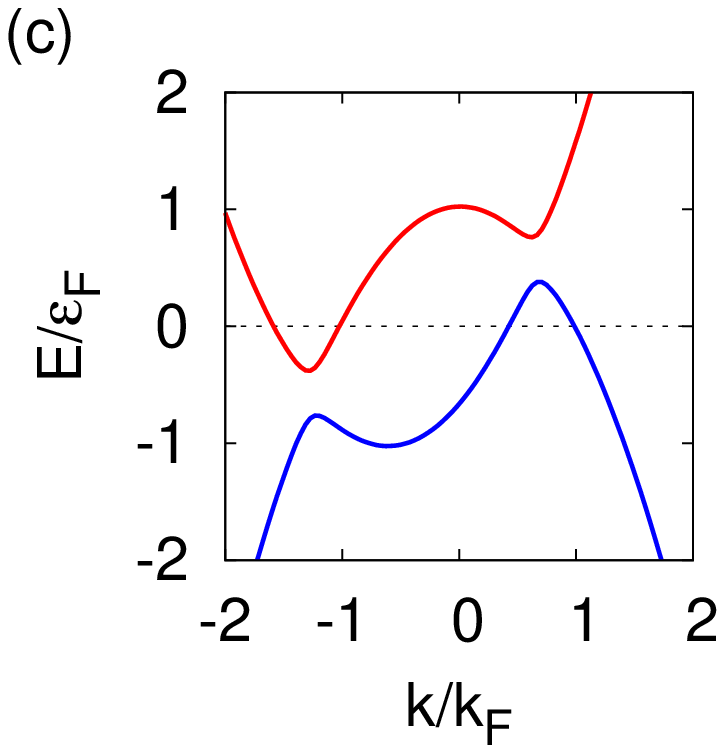}
\end{center}
\end{minipage}
\hspace{0.5cm}
\begin{minipage}{0.35\hsize}
\begin{center}
\includegraphics[width=\hsize,height=\hsize]{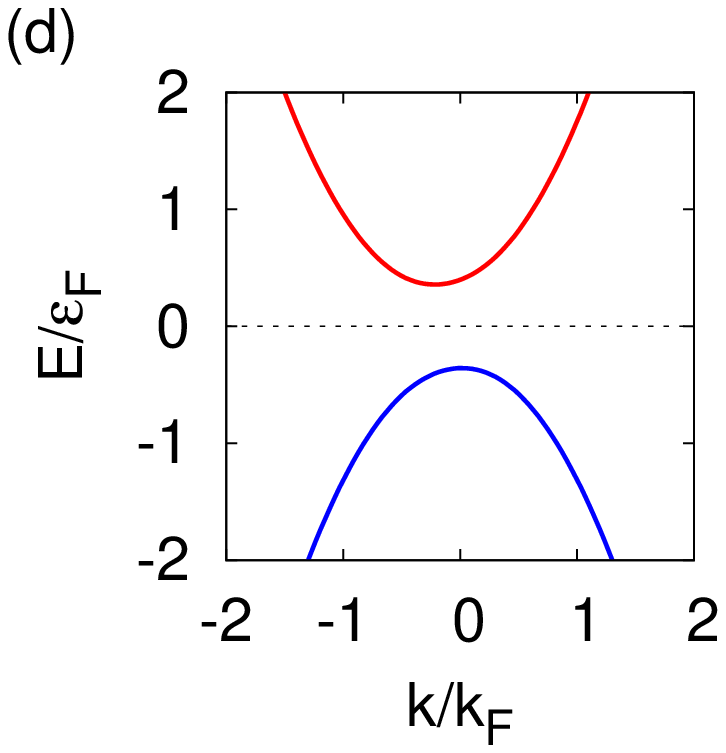}
\end{center}
\end{minipage}
\end{tabular}
\caption{
Dispersions in the BCS states in one-dimension at
$\mu=\varepsilon_F, \Delta_0=0.2\varepsilon_F$ for 
(a) $v_s=0$, (b) $v_s=v_L$, and (c) $v_s=3v_L$,
where $v_s=|\vecc{q}|/2m_0$ and
$v_L=\Delta_0/k_F$. 
(d) Dispersion in the BEC state at 
$\mu=-0.3\varepsilon_F, \Delta_0=0.2\varepsilon_F,
v_s=v_L$.
Red curve is $E_{k1}$ and blue curve is $E_{k2}.$}
\label{fig:FF}
\end{figure}

We consider a non-interacting $s$-wave superfluid with a modulating gap function $\Delta_0 \exp (i\vecc{q}\vecc{r})$ 
under periodic boundary conditions,
\begin{align}
{H}_{\rm FF}&=\sum_{k\sigma}
\varepsilon_kc^{\dagger}_{k\sigma}c_{k\sigma}+\sum_k\Delta_0 
c^{\dagger}_{k+q\uparrow}c^{\dagger}_{-k\downarrow}+{\rm (h.c.)}\notag\\
&=\sum_k
\left[
\begin{array}{c}
c^{\dagger}_{k+q\uparrow}\\
c_{-k\downarrow}
\end{array}\right]^T
\left[
\begin{array}{cc}
\varepsilon_{k+q}&\Delta_0\\
\Delta_0&-\varepsilon_{-k}
\end{array}\right]
\left[
\begin{array}{c}
c_{k+q\uparrow}\\
c^{\dagger}_{-k\downarrow}
\end{array}\right]
\label{eq:HamFF}
\end{align}
where $\varepsilon_k=k^2/2m-\mu$.
The Hamiltonian has a conserved quantity
\begin{align}
{\vecc{\mathcal P}}={\vecc{P}}-\vecc{q}{N}/2,
\end{align}
where ${N}=\sum_{k\sigma}c_{k\sigma}^{\dagger}c_{k\sigma}$ and 
${\vecc{P}}$ is the total linear momentum
${\vecc{P}}=\sum_{k\sigma}\vecc{k}c_{k\sigma}^{\dagger}c_{k\sigma}$.
This momentum $\vecc{\mathcal P}$ characterizes the deviation from the naively expected value $\langle \vecc{P}\rangle_V=\vecc{q}\langle N\rangle_V /2$.
Eigenvalues of the BdG Hamiltonian $H_k$ are
\begin{align}
E_{k1,2}&=\frac{1}{2}\Bigl[\varepsilon_{k+q}-\varepsilon_{-k}
\pm\sqrt{(\varepsilon_{k+q}+\varepsilon_{-k})^2+4\Delta_0^2}\Bigr].
\end{align}
$E_{ki}$ vanishes when $|\varepsilon_{k+q}-\varepsilon_{-k}|=
\sqrt{(\varepsilon_{k+q}+\varepsilon_{-k})^2+4\Delta_0^2}$ or
equivalently $\varepsilon_{k+q}\varepsilon_{-k}+4\Delta_0^2=0$
which can be satisfied only in the BCS regime $\mu>0$.
The momentum space is divided into three regions depending on
signs of the eigenvalues, 
$K_1=\{\vecc{k}|E_{k1}<0, E_{k2}<0\}$,
$K_2=\{\vecc{k}|E_{k1}>0, E_{k2}>0\}$, and
$K_3=\{\vecc{k}|E_{k1}>0, E_{k2}<0\}$
as shown in Fig. \ref{fig:FF}. 
Note that the volumes of $K_{1,2}$ are the same by the particle-hole symmetry.
Then, the ground state wavefunction determined by the conditions $b_{ki}|{\rm GS}\rangle=0 (E_{ki}>0)$
and $b_{ki}^{\dagger}|{\rm GS}\rangle=0 (E_{ki}<0)$,
and is given by
\begin{align}
|{\rm GS}\rangle &
={\mathcal N}\left(\prod_{k\in K_1}c^{\dagger}_{k+q\uparrow}\right)
\left(\prod_{k\in K_2}c^{\dagger}_{-k\downarrow}\right)\notag\\
&\quad\times 
\prod_{k\in K_3}\exp\left(-v_k/u_kc^{\dagger}_{k+q\uparrow}c^{\dagger}_{-k\downarrow}\right)
|0\rangle,
\label{eq:FF}
\end{align}
where $(u_k,v_k)^T$ are eigenvectors of the Bogoliubov-de Gennes
equation and ${\mathcal N}$ is a normalization constant.
One can find an essential similarity between Eqs.~\eqref{eq:FF} and \eqref{eq:GS}.
Note that, for example when $q_x>0$, $k_x+q_x<0$ for $\vecc{k}\in K_1$ is satisfied,
and breaking of a Cooper pair $c_{k+q\uparrow}^{\dagger}c_{-k\downarrow}^{\dagger}|0\rangle
\rightarrow c_{k+q\uparrow}^{\dagger}|0\rangle$ will reduce the linear momentum as
$(k_x+q_x)+(-k_x)=q_x>0\rightarrow (k_x+q_x)<0$ as in the OAM of neutral superfluids discussed in the previous section.
Similar reduction of the linear momentum takes place for $\vecc{k}\in K_2.$

The reduction of the linear momentum can be clearly discussed based on the unpaired fermions and fermionic Landau criterion.
The expectation value of the momentum deviation is 
$\langle {\rm GS}|
{\vecc{\mathcal P}}|{\rm GS}\rangle=-1/2\sum_k (\vecc{k}+\vecc{q}/2)
\sum_i {\rm sgn}E_{ki}$, or equivalently,
\begin{align}
\langle \vecc{P}\rangle_V&=\vecc{q}\frac{\langle N\rangle_V}{2}-\frac{1}{2}\sum_k(\vecc{k}+\vecc{q}/2)\eta_k,\\
\eta_k&=\sum_{i=1,2}{\rm sgn}E_{ki}.
\end{align}
If the supercurrent
velocity $v_s=|\vecc{q}|/2m_0$ is smaller than the Landau critical velocity 
$v_L=\Delta_0/k_F$, 
$K_{1,2}$ are empty and  
functional structure of $|{\rm GS}\rangle$ is essentially same as
that of the conventional full gap BCS state and $\langle \vecc{\mathcal P}\rangle_V=0$, i.e. $\langle\vecc{P}\rangle_V=\vecc{q}\langle N\rangle_V/2$.
On the other hand, 
when the supercurrent velocity is sufficiently fast, $v_s\gtrsim v_L$,
the two regions $K_{1,2}$ are non-empty, leading to 
$|\langle\vecc{P}\rangle_V|<|\vecc{q}|\langle N\rangle_V /2$ in the BCS regime where $\mu>0$.
The ground state now contains a non-zero fraction of normal state fermions, i.e. unpaired fermions,
and the system has both a Fermi surface and Cooper pairs.
This means that, 
the fast flow causes depairing of the Cooper pairs for the fermions with
$|\varepsilon_{k+q}-\varepsilon_{-k}|\gtrsim2\Delta_0$.
Therefore, the present toy model is quite analogous to the original setup of the bosonic Landau criterion.
It is noted that,
in contrast to the BCS regime with $\mu>0$,
$K_{1,2}$ are always empty and $\langle\vecc{P}\rangle_V=\vecc{q}\langle N\rangle_V/2$ in the BEC regime
where $\mu<0$ as shown in Fig. \ref{fig:FF} (d).

As was mentioned in the previous section,
the Landau criterion does not necessarily hold in general interacting models.
In an interacting model of the Fulde-Ferrell superfluid,
a large $q>v_L$ will eventually destroy the whole superfluidity and the system will become a normal state (non-superfluid)~\cite{book:Tinkham2004,pap:Bardeen1962}.
To evaluate the stability of the pre-assumed gap function, we need to calculate 
the ground state energy or free energy of the interacting model.
This is also true for other non-trivial state with unpaired fermions, such as the breached pair state 
where gapless modes with a Fermi surface coexist with paired fermions~\cite{pap:breached1,pap:breached2}.

\section{iDMRG calculation of mass current}
\label{sec:DMRG}
In the previous section, we have established the physical picture of the fragile spontaneous OAM in neutral fermionic superfluids
based on the mean field approximation.
A natural question is that whether or not the physics within the mean field description can be justified 
when we fully include interactions.
In this section, we try to go beyond the mean field approximation by treating many-body interactions properly.

Here, we consider a model of the chiral $p$-wave superfluid with domains of opposite superfluid chiralities for an illustrative perpose.
Within the mean field approximation, it is known that domain wall current is reversed depending on details of
the domain boundary in such a system~\cite{pap:Tsutsumi2014,pap:Tada2018}; the domain wall mass current flows in a certain direction for the $(p_x+ip_y)/(p_x-ip_y)$ domain junction,
while it is in an opposite direction for the $(p_x+ip_y)/(-p_x+ip_y)$ domain junction, which can be understood in terms of
the unpaired fermions and fermionic Landau criterion~\cite{pap:Tada2018}.
This is a drastic change in the domain wall mass current,
and it would be relatively easy to discuss whether this holds true beyond the mean field approximation.
It is noted that the domain wall mass current and edge mass current have essentially the same origin in common,
and both of their dependences on boundaries can be understood based on the unpaired fermions and fermionic Landau criterion in the same way~\cite{pap:Tada2015PRL,pap:Tada2018}.
Therefore we expect that studying the former is relevant to the latter.

\begin{figure}[htbp]
\vspace{25pt}
\begin{center}
\includegraphics[width=0.9\hsize,height=0.3\hsize]{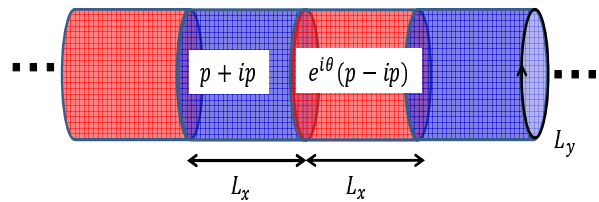}
\end{center}
\caption{Infinite length cylinder with the circumference $L_y$.
Each domain size is $L_x\times L_y$.}
\label{fig:cylinder}
\end{figure}

Our Hamiltonian is a spinless fermion model with nearest neighbor attractive interaction on a two-dimensional square lattice,
\begin{align}
H&=\sum_{i,j}-t_{ij}c_{i}^{\dagger}c_{j}+V_0\sum_{\langle i,j\rangle}n_in_j+H_{\rm SB},\\
H_{\rm SB}&=\sum_{\langle i,j\rangle}[\Delta^0_{ij}c_{i}^{\dagger}c_{j}^{\dagger}+({\rm h.c.})],
\end{align}
where $t_{ij}=t (i\neq j)$ is the nearest neighbor hopping and $t_{ii}=\mu$ is the chemical potential.
We have introduced a small symmetry breaking field of U(1) symmetry $\Delta^0_{ij}$
which describes a domain structure.
Although the symmetry breaking field must be $\Delta^0\rightarrow0$ after the thermodynamic limit is taken,
we keep a small value of $\Delta^0_{ij}$ and discuss effects of the interaction $V_0$ in a finite size system.
This calculation allows us to examine whether or not the mean field understanding is essentially correct.
We apply infinite density matrix renormalization group (iDMRG) 
and use the open source code TenPy~\cite{pap:DMRG0,pap:DMRG1,pap:DMRG2,pap:DMRG3,pap:TenPy,pap:TenPy2}.
The system size is $\infty\times L_y$ with periodic boundary
condition for the $y$-direction.
To realize the domain structure shown in Fig.~\ref{fig:cylinder}, 
the symmetry breaking field is taken to be $\Delta^0_{ij}=e^{i\theta_i}\Delta_0(\delta_{i,j+\hat{x}}\pm i\delta_{i,j+\hat{y}})$ for a site
$i=(x,y)$ when $nL_x\leq x<(n+1)L_x$ and $(n+1)L_x\leq x<(n+2)L_x$ respectively,
where $L_x$ is the domain size and $n$ is an integer. 

The phase $\theta_i$ characterizes the structure of a domain wall;
(I) $\theta_i=0$ for which $\Delta_{i,j+\hat{y}}$ changes the sign at the boundary
and (II) $\theta_i=\delta_{x,nL_x}\pi$ for which $\Delta_{i,j+\hat{x}}$ changes the sign.
The chirality of a domain is independent of $\theta_i$.
It is known that the domain wall corresponding to (I) is more stable than that of (II) within mean field calculations~\cite{pap:Tada2018,pap:SalomaaVolovik1989}.
The important point is that the directions of the domain wall currents for the two domain walls
are opposite, which is rather counter-intuitive since a domain wall current in a chiral $p$-wave state is usually determined by the chirality of the gap function.
This non-trivial behavior can be understood based on the unpaired fermions and fermionic Landau criterion~\cite{pap:Tada2018}.
Here, we discuss validity of the physical understanding based on the mean field approximations 
with use of iDMRG which are essentially free from approximations~\cite{pap:DMRG0,pap:DMRG1,pap:DMRG2,pap:DMRG3,pap:TenPy,pap:TenPy2}.

Now we numerically evaluate the mass current density for each domain wall, $\theta=0$ or $\theta=\pi$.
We have done similar calculations for several values of the symmetry breaking field $\Delta_0$ and
different system sizes $L_x, L_y$, and they show qualitatively similar results.
In the following, we focus on the smallest symmetry breaking field $\Delta_0=0.01t$ used in the calculations, and fix the system size as $L_x=L_y=8$.
The bond dimension $\chi$ controls accuracy of the iDMRG calculations, 
and we used only three values $\chi=200,400,600$. Although these are not sufficient to obtain fully convergent results,
they give qualitatively same results. 
Therefore, we fix $\chi=400$ in the following
to discuss the validity of the mean field approximations, for which the truncation norm error is $O(10^{-4})$.
The relatively small truncation error for the parameters used is due to the symmetry breaking field $\Delta_0$ which makes the bulk of the system
gapped.

\begin{figure}[htbp]
\begin{tabular}{cc}
\begin{minipage}{0.5\hsize}
\begin{center}
\includegraphics[width=\hsize,height=0.6\hsize]{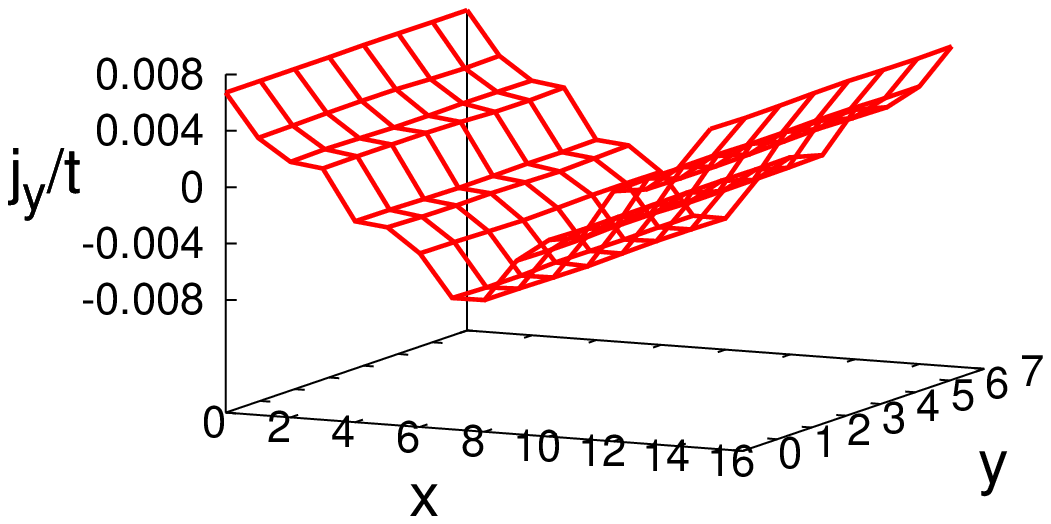}
\end{center}
\end{minipage}
\begin{minipage}{0.5\hsize}
\begin{center}
\includegraphics[width=\hsize,height=0.6\hsize]{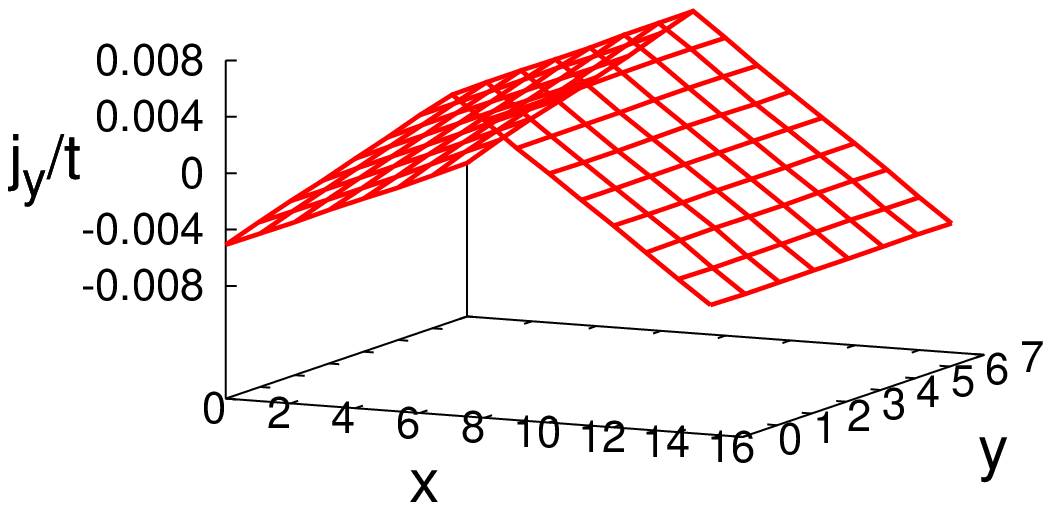}
\end{center}
\end{minipage}
\end{tabular}
\caption{Mass current density along the $y$-direction for $\theta=0$ (left) and $\theta=\pi$ (right).
The parameters are $V_0=-0.1t, \mu=-t$.
The chirality of $\Delta^0$ is positive for $0\leq x<L_x=8$, while it is negative for $L_x\leq x<2L_x=16$.
}
\label{fig:jyV01}
\end{figure} 
\begin{figure}[htbp]
\begin{tabular}{cc}
\begin{minipage}{0.5\hsize}
\begin{center}
\includegraphics[width=\hsize,height=0.6\hsize]{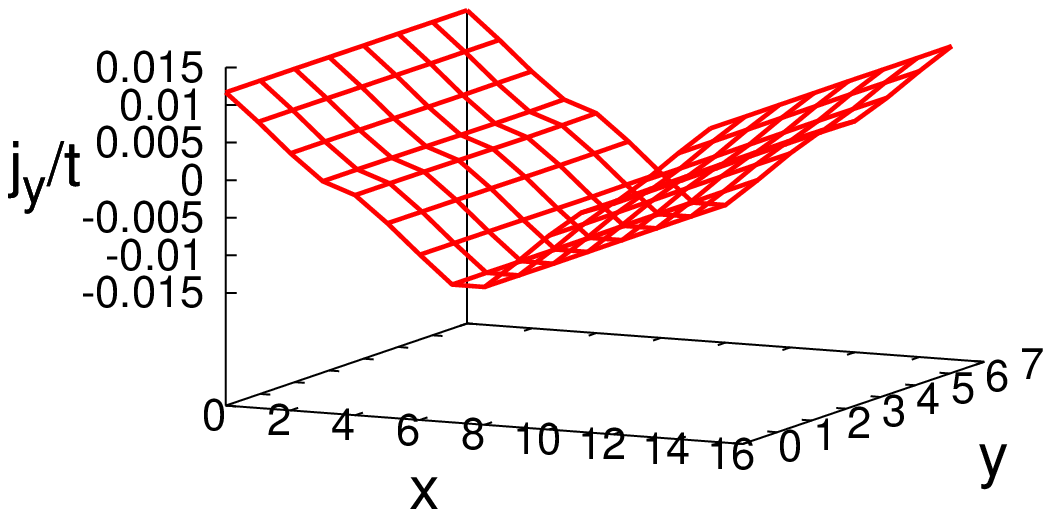}
\end{center}
\end{minipage}
\begin{minipage}{0.5\hsize}
\begin{center}
\includegraphics[width=\hsize,height=0.6\hsize]{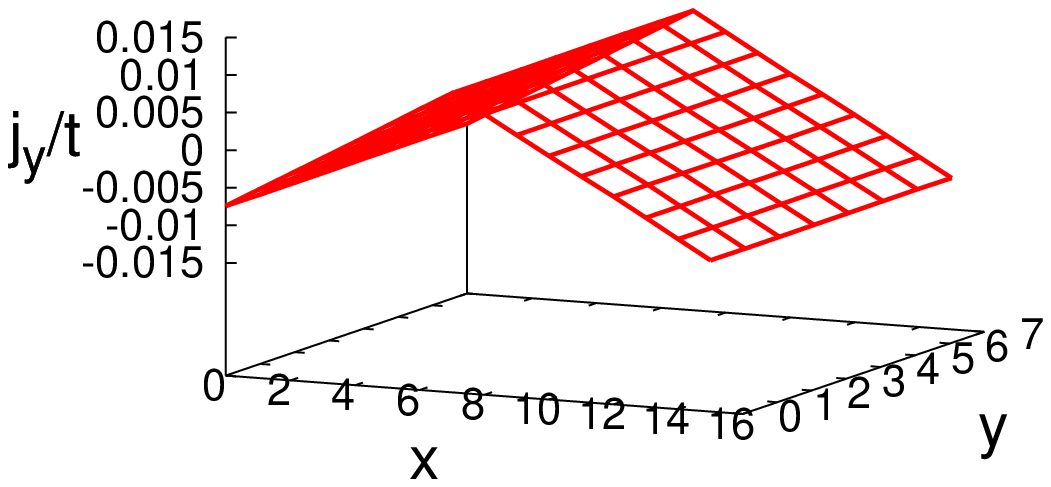}
\end{center}
\end{minipage}
\end{tabular}
\caption{Mass current density along the $y$-direction for $\theta=0$ (left) and $\theta=\pi$ (right).
The parameters are $V_0=-t, \mu=-2.2t$.
The chirality of $\Delta^0$ is positive for $0\leq x<L_x=8$, while it is negative for $L_x\leq x<2L_x=16$.
}
\label{fig:jyV1}
\end{figure} 

We show in Fig.~\ref{fig:jyV01} the calculated mass current densities at 
a small interaction $V_0=-0.1t, \mu=-t$ for which the average particle filling is $n_{\rm }\simeq 0.32$.
Unfortunately, the current profile does not show sharp localization at a domain boundary, since
the system size $L_x=L_y=8$ is not large enough compared with the coherence length for the parameters used.
Instead, the current profile shows a broad structure where $|j_y|$ is largest at the domain boundary, while it is smallest in the middle 
of a domain.
Nevertheless, the current directions for $\theta=0$ and $\theta=\pi$ are opposite, which is consistent with the mean field calculations~\cite{pap:Tsutsumi2014,pap:Tada2018}.
The ground state energy difference between the two states is small, $[E(\theta=0)-E(\theta=\pi)]/[E(\theta=0)+E(\theta=\pi)]=O(10^{-3})$.
Now we increase the interaction and find that
the current reversal is stable even for a relatively large interaction $V_0=-t, \mu=-2.2t$ for which
$n_{\rm }\simeq 0.28$, as shown in Fig. \ref{fig:jyV1}.
The mass current is enhanced by the interaction $V_0$ and 
the maximum density becomes nearly double compared with those for $V_0=-0.1t$ if $j_y$ is normalized by the filling $n_{\rm}$.
This means that $j_y$ is dominated by the interaction $V_0=-t$ and the current reversal gets stabilized.
The origin of the enhancement of $j_y$ would be the decreased superfluid coherence length $\xi_{\Delta}\simeq t/\Delta$ by the interaction $V_0$
where $\Delta$ is the gap amplitude,
because a small $\xi_{\Delta}$ gives $j_y$ which is well localized around a boundary and does not influence $j_y$ at the opposite boundary.
If we increase $|V_0|$ further, the simulations become unstable and the fermions get dimerized.
These numerical results suggest that the current reversal found in the mean field calculations holds true even in the
iDMRG calculations which are essentially free from approximations.
Therefore, we conclude that the physical understanding based on the mean field approximation is essentially correct,
and the physics is determined by the unpaired fermions and fermionic Landau criterion.
Finally, although our iDMRG results support the correctness of the mean field understanding, 
they are reliable at a rather qualitative level and further numerical calculations would be required to develop a quantitative understanding.

\section{summary and discussion}
\label{sec:summary}
In this study, we have discussed the OAM and corresponding edge or domain wall mass current in neutral fermion superfluids with broken time reversal symmetry.
It was explained that OAM in a neutral superfluid cannot be obtained by derivative of the thermodynamic free energy with respect to 
its intensive conjugate external field in sharp contrast to non-superfluid systems.
This means that OAM in a neutral superfluid is not a thermodynamic quantity and can be influenced by non-thermodynamic details.
We established a simple physical picture of how OAM is changed by such perturbations based on the mean field approximation,
by introducing the concepts of the unpaired fermions and fermionic Landau criterion.
We also discussed the validity of the mean field description by a non-perturbative numerical calculation using iDMRG.
It is concluded that the mean field calculations of OAM and edge mass current for chiral superfluids 
are essentially correct, and OAM does depend on non-thermodynamic details such as boundary conditions which are usually not controlable in experiments.
The sensitivity of OAM can be considered as an anomalously colossal response of OAM to boundaries.
If one could control the boundary conditions of a neutral superfluid, 
a dramatic response of OAM by small perturbations might be obtained.

In the original problem of the ``intrinsic angular momentum paradox" in $^3$He-A phase in three dimensions,
the rotation axis of Cooper pairs will locally deviates near the wall of a container from that in the bulk~\cite{book:Vollhart1990,book:Volovik2003,
pap:Leggett1975,book:Leggett2006,lecture:Leggett2012,pap:Mizushima2016}.
Because this effect may depend on the container used, some people have anticipated that this problem would depend
on theoretical models used and experimental details.
However, such a subtle problem is absent 
in simpler systems such as the thin film limit of a chiral $p$-wave state and two-,three-dimensional chiral $d$-wave state,
and one may expect an ``intrinsic" value of the spontaneous OAM.
What we have discussed in this study is that, even in such a relatively simple system,
there is no ``intrinsic" value of OAM in a uniform neutral superfluid,
simply because it is not a thermodynamic quantity.

The OAM and edge mass current of a neutral superfluid will be determined for each given surface condition and sample shape.
There exist various possible perturbations such as surface adsorption, surface reconstruction, and surface disorder,
and surface conditions should be carefully treated in experiments, although it is a very difficult issue.
Sample shapes and relative relations between the surface direction and underlying lattice geometry should also be controlled in solid state superconductors,
to observe edge charge currents.

Finally, we touch on charged chiral superconductors such as the candidate $p$-wave superconductor Sr$_2$RuO$_4$. 
It is considered that the edge current depends on boundaries in this system~\cite{ pap:Hicks2010,pap:Scaffidi2015}. 
In such a system, there arises a Meissner screening current in addition to a spontaneous edge current. 
The former is localized at a surface in the length scale of coherence length, while the latter is in the length scale of penetration depth, 
and they are spatially separated. 
These two contributions will cancel each other in a longer length scale, and the total net edge current $\vecc{J}_{\rm edge}$ 
and corresponding spontaneous OM vanishes in absence of an external field~\cite{pap:Tada2018}. Since the spontaneous edge current 
which we have discussed in the present study depends on boundaries and shapes of the system, the corresponding screening current also depends on them. 
Therefore, the induced local magnetic flux density which is to be measured in experiments would also be sensitive to boundaries and shapes.

\section*{acknowledgement}
We are grateful to F. Pollmann for introducing the open source code TenPy to us and the helpful comments on our manuscript.
We also thank T. Koma, M. Oshikawa, S. Sugiura, H. Tsunetsugu, T. Ikeda, O. Sugino, H. Akai, 
S. Fujimoto, T. Mizushima,
Y. Fuji, Y. Wan, and P. Fulde for valuable discussions.
This work was supported by JSPS/MEXT Grant-in-Aid for Scientific Research
(Grant No. 26800177 and No. 17K14333)
and by a Grant-in-Aid for
Program for Advancing Strategic International Networks to
Accelerate the Circulation of Talented Researchers (Grant No.
R2604) ``TopoNet.''

\appendix
\section{Thermodynamic limit under uniform flux density}
\label{app:A}
We explain that the thermodynamic limit exists for a U(1) symmetric system with a given uniform magnetic flux density.
The existence proof is the same as in the previous studies, once one notices that the Hamiltonian is stable and translationally symmetric
with an appropriate gauge transformation~\cite{book:Ruelle1999}.
Nevertheless, as a reference, here we will give a brief discussion on both lattice models and continuum models with stable, short-range interactions.
It should be noted that existence of a thermodynamic limit for a system with long-range interactions is highly non-trivial,
and for example, one would find the familiar shape/boundary condition dependent free energy density of a magnet with dipole interactions
in the presence of an external field
~\cite{pap:Banerjee1998,pap:Campa2009}.
Long range interactions or dynamics of electromagnetic field is beyond the present study.
We also discuss Bloch's theorem on absence of a macroscopic current at equilibrium as a collorary of the existence of the thermodynamic limit.

\subsection{Lattice model}
We consider a simple model defined on a lattice $\Lambda\subset {\mathbb Z}^d$ where $d=3$ is the system dimension,
\begin{align}
H_{\Lambda}&=\sum_{i,j}-t_{ij}c_{i}^{\dagger}c_{j}+V\sum_{\langle i,j\rangle}n_in_j,\notag \\
&\equiv \sum_{X\subset\Lambda}h_X,
\end{align}
where $X=\{i\},\langle i,j\rangle$ represents sites or nearest neighbor pairs of sites.
The hopping term contains a given vector potential $A_{ij}$ which realizes a uniform magnetic flux density along $z$-axis,
and its amplitude $|t_{ij}|$ is constant. We have also included the chemical potential, $t_{ii}=\mu$.
It is important to see that the Hamiltonian is symmetric under the magnetic translation. 
We denote the $n_{\mu}$-sites magnetic translation operator along $\mu$-direction as $T_{\mu}(n_{\mu})$.
Then, for a translation $T(n)=T_x(n_x)T_y(n_y)T_z(n_z)$,
\begin{align}
H_{\Lambda+n}=T(n)H_{\Lambda}T(n)^{-1},
\end{align}
where $\Lambda+n$ is the translate of $\Lambda$ by the vector $n$.
This relation is independent of the order of $T_x,T_y,T_z$ in $T$, since they are a projective representation of translation.
Therefore, the free energy density $f_{\Lambda}=F_{\Lambda}/|\Lambda|$ is also translationally symmetric
\begin{align}
f_{\Lambda+n}=f_{\Lambda}.
\end{align}

For simplicity, we consider a cube $\Lambda_a=\{x\in {\mathbb Z}^d|0\leq x_{\mu}< a\}$
and a larger cube 
$\Lambda=\{x\in {\mathbb Z}^d|0\leq x_{\mu}< la\}$ where $l,a$ are positive integers.
We define $\Lambda_{a+n}=\{x\in {\mathbb Z}^d|n_{\mu}a\leq x_{\mu}< (n_{\mu}+1)a\}$ and denote them as $\{\Lambda_j\}_{j=1}^{l^d}$
so that $\Lambda=\cup_{j=1}^{l^d} \Lambda_{j}$.

We compare the free energy densities $f_{\Lambda}$ and $f_{\Lambda_a}$.
For $\Gamma_j=\cup_{i=1}^{j}\Lambda_i$, one can show
\begin{align}
|F_{\Lambda}-\sum_{j=1}^{l^d}F_{\Lambda_j}|\leq ||h|| \sum_{j=2}^{l^d}N(\Gamma_{j-1},\Lambda_j),
\end{align}
where $N(\Gamma_{j-1},\Lambda_j)$ is the number of sites for which $h_{\langle k,l\rangle}\neq0$ with 
$k\in \Gamma_{j-1}$ and $l\in \Lambda_j$.
We have introduced $||h||=\sum_{X\ni 0}||h_X||/|X|$ which is well defined because $h_X$ is magnetic translationally symmetric.
It is important to see that the seemingly dangerous terms in $h_X$ close to the boundary of $\Lambda_j$ are harmless in the presence of the vector potential.
Since $N(\Gamma_{j-1},\Lambda_j)=O(a^2)$ for large $a$,
we have
\begin{align}
|f_{\Lambda}-f_{\Lambda_a}|= O(1/a),
\end{align}
which means the existence of the thermodynamic limit $f_{\infty}=\lim_{\Lambda\rightarrow {\mathbb Z}^d}f_{\Lambda}$.
One can also show the existence of the thermodynamic limit for a more general sequence of lattices,
and the resulting $f_{\infty}$ is independent of the system shape for such lattices.

\subsection{Continuum model}
We next touch on continuum models.
Although discussions on continuum models are generally complicated, 
the proof for Hamiltonian with usual translation symmetry can also be applied to that
with magnetic translational symmetry~\cite{book:Ruelle1999}.

We consider a Hamiltonian of $N$ fermions,
\begin{align}
H_{\Lambda,N}&=T_{\Lambda,N}+U_{\Lambda,N},\\
T_{\Lambda,N}&=\sum_{j=1}^N\frac{1}{2m}\bigl(-i\nabla_j-\vecc{A}(\vecc{r}_j)\bigr)^2,\\
U_{\Lambda,N}&=\sum_{i<j}U(|\vecc{x}_i-\vecc{x}_j|), 
\end{align}
where the vector potential gives a constant magnetic flux density. 
The interaction is stable ($U_{\Lambda,N}>-bN$ with $b\geq0$)
and short-range with the range $r_0$ or strongly tempered, $U(r> r_0)\leq0$. 
$\Lambda\subset {\mathbb R}^d$ is a bounded region, and we consider wavefunctions which smoothly tend to zero at the boundary $\partial \Lambda$
and vanishes outside of $\Lambda$.
Mathematically, $H_{\Lambda,N}$ should be regarded as a self-adjoint Friedrichs extension.

The spectrum of $H_{\Lambda,N}$ consists of discrete eigenvalues with finite multiplicity, and we denote them
in increasing order as $E_1\leq E_2\leq \cdots$.
Then, the minimax principle reads
\begin{align}
E_m=\inf_{{\mathcal M}:{\rm dim}{\mathcal M}=m}\sup_{\phi:\phi\in {\mathcal M},||\phi||=1}\langle \phi| H|\phi\rangle.
\end{align}
Now we discuss entropy $S_{\Lambda}(N,E)=\log W_{\Lambda}(N,E)$, where $W_{\Lambda}(N,E)$ is the
number of eigenvalues of $H_{\Lambda,N}$ below $E$.
Note that $S_{\Lambda}$ is translationally invariant, since the Hamiltonian has magnetic translational symmetry.
We consider two regions $\Lambda_i (i=1,2)$ which are separated by the distance $r\geq r_0$,
and $N_i$ particles are confined in each region, respectively.
The states with energy of $H_{\Lambda_i,N_i}$ below $E_i$ are described by wave functions $\varphi_i\in {\mathcal M}_i$ which satisfy the hard wall boundary condition
$\varphi_{i}(x\in \partial\Lambda_i^{N_i})=0$, for which dim${\mathcal M}_i=\Omega_{\Lambda_i}(N_i,E_i)$.
From the minimax principle, we have
\begin{align}
\sup_{\varphi_i\in {\mathcal M}_i,||\varphi_i||=1}\langle \varphi_i| H_{\Lambda_i,N_i}|\varphi_i\rangle\leq E_i.
\end{align}
Now we construct a subspace ${\mathcal M}_{1+2}$ of the total Hilbert space for particles in $\Lambda_1\cup\Lambda_2$;
${\mathcal M}_{1+2}$ is generated by antisymmetrized $\varphi_1\otimes \varphi_2$ and its dimension is 
dim${\mathcal M}_{1+2}=\Omega_{\Lambda_1}(N_1,E_1)\cdot \Omega_{\Lambda_2}(N_2,E_2).$
Note that $\varphi\in {\mathcal M}_{1+2}$ vanishes on $(\partial\Lambda_1)\cup\Lambda_2$ or $\Lambda_1\cup(\partial\Lambda_2)$,
and seemingly dangerous contributions to $\Omega_{\Lambda_1\cup\Lambda_2}(N_1+N_2,E)$ from those boundaries are harmless
in the presence of the vector potential $|\vecc{A}|\propto |\vecc{r}|$, as in lattice models.
Since $\sup_{\varphi\in {\mathcal M}_{1+2},||\varphi||=1}\langle \varphi| H_{\Lambda_1\cup\Lambda_2,N_1+N_2}|\varphi\rangle
\leq E_1+E_2$,
we obtain
\begin{align}
S_{\Lambda_1\cup\Lambda_2}(N_1+N_2,E_1+E_2)\geq S_{\Lambda_1}(N_1,E_1)+ S_{\Lambda_2}(N_2,E_2).
\label{eq:ineq_S}
\end{align}
Furthermore, $S_{\Lambda}$ is an increasing function of $\Lambda$, i.e. $S_{\Lambda}\leq S_{\Lambda'}$ if $\Lambda\subset \Lambda'$.
Therefore, for a sequence of regions $\Lambda_j=\{x\in{\mathbb R}^d|0\leq x\leq L_j=2L_{j-1}+r_0\}$ such that 
$\Lambda_{j}$ contains $2^d$ translates of $\Lambda_{j-1}$ with mutual distance $r_0$,
$s_{\Lambda_j}=S_{\Lambda_j}(N_j,E_j)/|\Lambda_j|$ with $N_j=2^dN_{j-1},E_j=2^dE_{j-1}$ is a non-decreasing sequence. 
Besides, entropy of the interacting model is bounded above by that of the corresponding non-interacting model,
\begin{align}
S_{\Lambda}(N,E)\leq S_{\Lambda}^{(0)}(N,E+bN),
\end{align}
since $E_m\geq E_m^{(0)}-bN$ holds, where $E_m^{(0)}$ is the $m$-th eigenvalue of $T_{\Lambda,N}$.
Therefore, there exists the thermodynamic limit of the entropy density, $s_{\infty}=\lim_{\Lambda\rightarrow{\mathbb R}^d} s_{\Lambda}$.
Because of the equivalence between different emsembles, thermodynamic limit of the free energy density also exists.

\subsection{Bloch's theorem}
We briefly discuss Bloch's theorem as a collorary of the existence of the thermodynamic limit of a U(1) symmetric system.
The theorem claims that a macroscopic current is not allowed at equilibrium~\cite{pap:TadaKoma2016}.
For simplicity, we consider a Hamiltonian $H$ of fermions with short-range interactions defined on 
a two-dimensional cylinder $\Lambda=S_{R}\times I$ where $S_R$ is a one-dimensional ring with radius $R$ and $I$ is an interval, 
under an external field $B_z$ perpendicular to $S_R$. 
There might possibly arise a uniform current density $\langle \vecc{j}(\vecc{r}\in \Lambda)\rangle=O(1)$ 
around the cylinder.
However, such a current density is not allowed, since if it exists,
the free energy will be superextensive, $|F_{\Lambda}(B_z\neq0)|=O(R^2\times |I|)\gg O(|\Lambda|)$
due to the coupling between $B_z$ and OM.
Therefore, there is no net current in $\Lambda$, which is
a variant of Bloch's theorem.

The above argument has a trivial but important physical implication.
Now we consider a three-dimensional ferromagnet which is fully wrapped with a thin film under the assumption that long range magnetic interactions are negligible.
The total system is a combination of the decoupled ferromagnet and thin film, and the latter is described as a two-dimensional system such as the cylinder in the above discussion.
Then, from the similar argument,
we conclude that it is impossible to change the value of OM of the ferromagnet by wrapping it with a thin film, which sounds rather trivial.
We can screen OM only when we use a superconducting thin film, where the Maxwell equation or long range magnetic interactions should be taken into account.
One can compare this with the electric polarization for which a constant electric field is not a conjugate intensive field~\cite{pap:Avron1977,pap:Blanc2002}.
Corresponding to the above argument, one can wrap a three-dimensional ferroelectric material by a two-dimensional metallic film, 
which can be regarded as a surface perturbation to the former.
Obviously, the polarization of the total system, i.e. the perturbed ferroelectric material, 
changes due to a screening effect by the thin metal. 
In sharp contrast to OAM/OM, the charge polarization can be easily affected by surface perturbations
even in a theoretical model with short range interactions.

\section{Numerical calculation of OAM}
\label{app:B}
\begin{figure}[htbp]
\begin{tabular}{cc}
\begin{minipage}{0.5\hsize}
\begin{center}
\includegraphics[width=0.8\hsize,height=0.6\hsize]{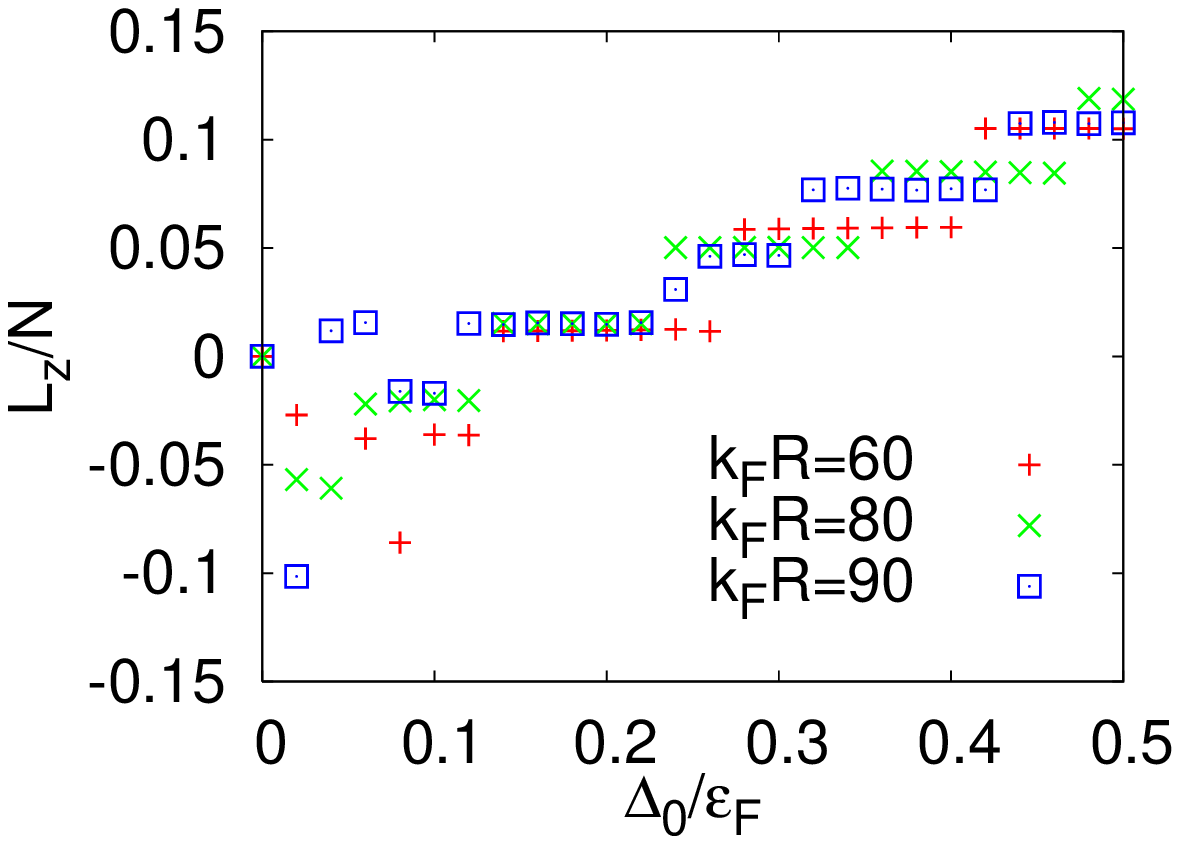}
\end{center}
\end{minipage}
\begin{minipage}{0.5\hsize}
\begin{center}
\includegraphics[width=0.8\hsize,height=0.6\hsize]{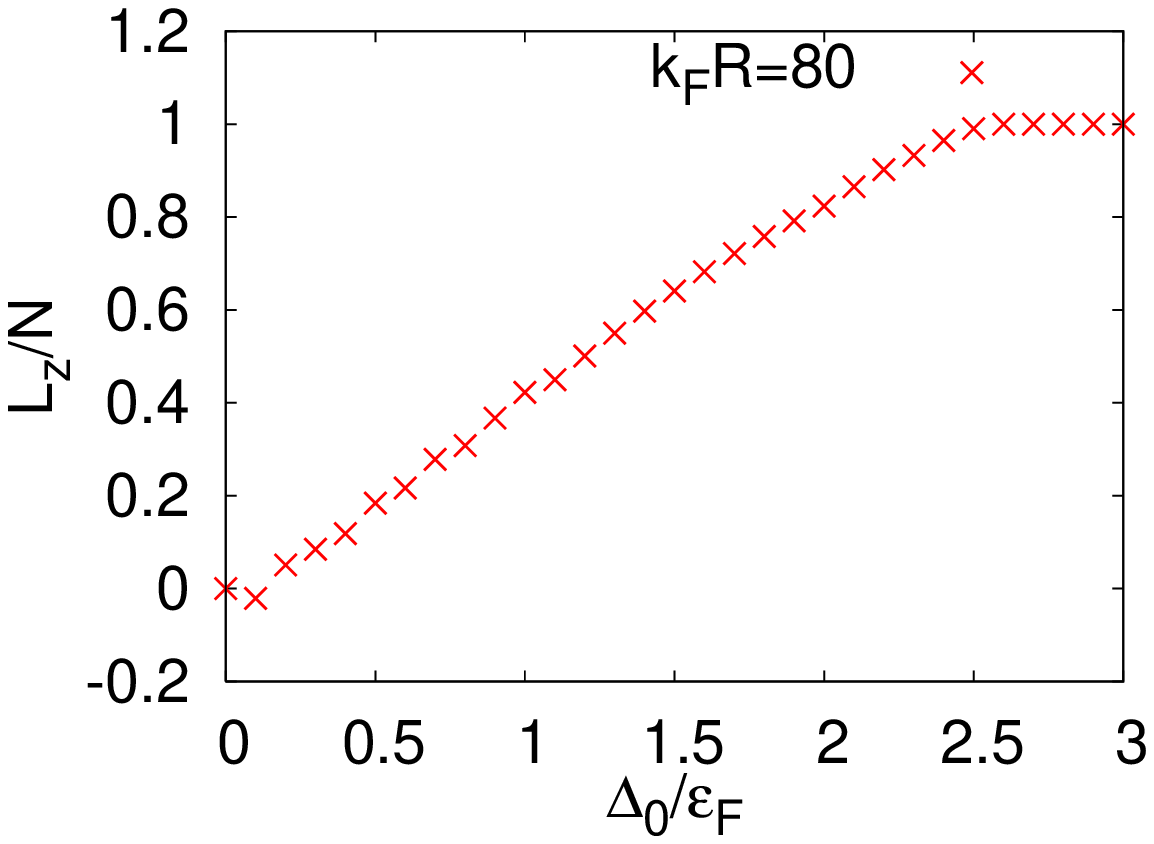}
\end{center}
\end{minipage}
\end{tabular}
\caption{OAM for the Hamiltonian Eq.~\eqref{eq:Hd}. $\Delta_0$ dependence for different system sizes (left panel),
and global $\Delta_0$ dependence for the fixed system size $k_FR=80$ (right panel).}
\label{fig:Lz_d}
\end{figure}

We breifly discuss numerical calculations of the OAM for the Hamiltonians Eq.~\eqref{eq:Hd}
with the sharp confinement potential, $V(r<R)=0, V(r>R)=\infty$. 
The calculation results are shown in Fig.~\ref{fig:Lz_d}.
The OAM per fermion is nearly zero for small $\Delta_0/\varepsilon_F$, although it is slightly oscillating around zero because of finite size effects.
The vanishing OAM is consistent with the semi-classical discussions given in the main text.
As $\Delta_0$ increases, the OAM changes discretely since the spectral asymmetry $\eta_l$ is an integer and the total number of fermions 
$\langle N\rangle_V$ is kept constant for each system size $k_FR$ with the same average density (see Eq.~\eqref{eq:Lz_eta}).
For large $\Delta_0$, the system enters the strong coupling BEC regime and the OAM takes the saturated value $\langle L_z\rangle_V/\langle N\rangle_V=1$~\cite{pap:Tada2015PRL}.


\begin{thebibliography}{98}
\bibitem [{com({\natexlab{a}})}]{com:OM_OAM}%
  \BibitemOpen
  \href@noop {} {} \bibinfo {note} {OM operator is defined
  as $M_z=(1/2c)\int_Vd^3x [\vecc{r}\times \vecc{j}_c]_z$ for a finite system
  with an open boundary condition, where $\vecc{j}_c=\vecc{j}_p+\vecc{j}_d$ is
  the gauge invariant charge current density given by the sum of the
  paramagnetic and diamagnetic currents. Since it is gauge invariant, we can
  choose $\vecc{A}=0$ for zero magnetic flux density $\vecc{B}=0$. In this
  particular gauge, $L_z=-\mu_BM_z$ holds as an operator identity in finite
  volume systems.}\BibitemShut {Stop}%
\bibitem [{com({\natexlab{d}})}]{com:F0}%
  \BibitemOpen
  \href@noop {} {} \bibinfo {note} {For a perturbation
  $\Delta H$, the change of the free energy is bounded as $\langle \Delta
  H\rangle_{V,H+\Delta H}\leq \Delta F_V\leq \langle \Delta H\rangle_{V,H}$
  from the Peierls inequality~\cite{book:Ruelle1999}, where $\langle
  \cdots\rangle_{V,H}$ is an expectation with the Hamiltonian $H$.}\BibitemShut
  {Stop}%
\bibitem [{com({\natexlab{e}})}]{com:extensive_operator}%
  \BibitemOpen
  \href@noop {} {} \bibinfo {note} {Here, we say that a
  bounded operator $M(V)$ defined on $V$ is extensive when $|M(V)|=O(V)$. For a
  unbounded operator we consider a subspace of the Hilbert space where $M(V)$
  is bounded.}\BibitemShut {Stop}%
\bibitem [{com({\natexlab{f}})}]{com:J_def}%
  \BibitemOpen
  \href@noop {} {} \bibinfo {note} {The net edge mass
  current $\vecc{J}_{\rm edge}$ is well-defined as long as $\vecc{j}_{\rm
  }(\vecc{r})$ is well-localized around the boundary of a system.}\BibitemShut
  {Stop}%
\bibitem [{com({\natexlab{j}})}]{com:Bq}%
  \BibitemOpen
  \href@noop {} {} \bibinfo {note} {The free energy density will diverge even when 
  one introduces an oscillating magnetic flux density such as $B_z^q(x)=B_z^0\cos(qx)$ 
  and take $q\rightarrow0$ limit. This is trivial when we take the $q\rightarrow0$ limit before 
  the thermodynamic limit $V\rightarrow\infty$ is taken to realize a $B$-field which is constant 
  over the whole system. We could also describe a nearly uniform $B$-field by taking 
  the $V\rightarrow\infty$ limit first, and then take $q\rightarrow0$. 
  In this case, for example, the induced current density will be essentially proportional 
  to the vector potential $A_y(x)=(B_z^0/q)\sin(qx)$ and it would diverge as $1/q$ for large $|x|$.
  }\BibitemShut {Stop}%
\end{thebibliography}

%

\end{document}